\def\mearth{{\rm\,M_\oplus}}

\documentclass[preprint,authoryear,12pt]{elsarticle}
\usepackage{amssymb}
\usepackage{aas_macros}

\begin{document}
 
\begin{frontmatter}

\title{Origin of water in the inner Solar System: planetesimals scattered inward during Jupiter and Saturn's rapid gas accretion}

\author{Sean N. Raymond$^{a}$ \& Andre Izidoro$^{a,b}$\\
\address[1]{Laboratoire d'Astrophysique de Bordeaux, Univ. Bordeaux, CNRS, B18N, all{\'e}e Geoffroy Saint-Hilaire, 33615 Pessac, France; rayray.sean@gmail.com}
\address[2]{UNESP, Univ. Estadual Paulista - Grupo de Din{\`a}mica Orbital
Planetologia, Guaratinguet{\`a}, CEP 12.516-410, S{\~a}o Paulo, Brazil; izidoro.costa@gmail.com}}

\begin{abstract}
There is a long-standing debate regarding the origin of the terrestrial planets' water as well as the hydrated C-type asteroids. Here we show that the inner Solar System's water is a simple byproduct of the giant planets' formation. Giant planet cores accrete gas slowly until the conditions are met for a rapid phase of runaway growth. As a gas giant's mass rapidly increases, the orbits of nearby planetesimals are destabilized and gravitationally scattered in all directions. Under the action of aerodynamic gas drag, a fraction of scattered planetesimals are deposited onto stable orbits interior to Jupiter's. This process is effective in populating the outer main belt with C-type asteroids that originated from a broad (5-20~AU-wide) region of the disk. As the disk starts to dissipate, scattered planetesimals reach sufficiently eccentric orbits to cross the terrestrial planet region and deliver water to the growing Earth. This mechanism does not depend strongly on the giant planets' orbital migration history and is generic: whenever a giant planet forms it invariably pollutes its inner planetary system with water-rich bodies. 
\end{abstract}

 

\end{frontmatter}

\section{Introduction}
The asteroid belt is a nearly empty expanse, with a total mass of only $\sim 5 \times 10^{-4} \mearth$~\citep{demeo13}. Asteroids have a radial compositional gradient: the inner belt is dominated by S-types and the outer belt by a variety of other classes but predominantly C-types~\citep{gradie82,bus02,demeo13,demeo14,demeo15}.  Asteroid classes have been spectroscopically associated with different types of meteorites~\citep{burbine02}. S-types are associated with ordinary chondrites and relatively dry, with $\le 0.1\%$ water by mass~\citep{robert77}.  C-types are linked with carbonaceous chondrites and are 5-20\% water by mass~\citep{kerridge85}.  

It is tempting to link the S- vs. C-type division with the water condensation line (the ``snow line'').  However, the snow line is not fixed but moves inward in time as the disk cools~\citep{lecar06,kennedy08,martin12}. The existence of ice at the snow line depends on inward drift of small particles~\citep{ciesla06}, which may be blocked by an outer, ice giant-mass object~\citep{lambrechts14b}, disconnecting the condensation temperature from the presence of water~\citep{morby16}. \cite{grimm93} argued that the timescale for planetesimal growth closer-in than 2.7 AU was short enough to capture energy from the short-lived radionuclide $^{26}$Al and to desiccate S-types, whereas C-types formed beyond 2.7 AU and more slowly, after $^{26}$Al was extinct. The Grand Tack model~\citep{walsh11} proposes that Jupiter formed beyond the snow line and that the asteroid belt was sculpted by the planet's inward-then-outward gas-driven migration. In this context, S-types represent planetesimals that originated interior to Jupiter and C-types planetesimals from between and beyond the giant planets' orbits that were dynamically implanted into the belt as a result of the giant planets' migration. 

Here we show that the giant planets' gas accretion destabilizes the orbits of nearby planetesimals, many of which are scattered inward. Some planetesimals are implanted into the outer asteroid belt, explaining the belt's compositional gradient. Others are scattered past the asteroid belt to deliver water to the growing terrestrial planets. This mechanism is an unavoidable side-effect of giant planet formation. Large-scale radial mixing of small bodies occurs naturally as planetesimals near the giant planets are destabilized. The existence and distribution of water in the inner Solar System is a simple consequence. The mechanism is robust to the giant planets' growth and migration history, and can improve a number of models of planet formation.

This paper is laid out as follows.  In section 2 we outline our simulations. In Section 3 we present results for the simple case of Jupiter and Saturn growing in-situ.   In Section 4 we show how the timing of giant planet formation determines the balance between asteroidal implantation and water delivery to the terrestrial planets.  In Section 5 we consider a range of possible growth and migration histories of the giant planets and explore the consequences. In Section 6 we present two simulations that also included the ice giants. In Section 7 we discuss the efficiency of asteroidal implantation and how it connects with planet formation scenarios.  In Section 8 we show how our results match Solar System constraints and how the mechanism can be incorporated within models of Solar System formation.  We conclude in Section 9. 

\section{Simulation methods}


We performed simulations designed to capture the state of a planet-forming disk mid-way through its evolution, after macroscopic bodies (planetesimals and giant planet cores) had formed but before cores had undergone runaway gas accretion.  Our code is built on the Symba/Swift integration package~\citep{levison94,duncan98}, modified to include a number of effects related to hydrodynamical interactions with an underlying gas disk.

Our simulations include one or two giant planet cores and a swarm of small planetesimals embedded in a gaseous disk. Jupiter and Saturn started as $3 \mearth$ cores -- just below the expected threshold for rapid gas accretion~\citep{ikoma01} -- and grew to their current sizes on a timescale of $\sim 10^5$ years~\citep[][; although we test the effect of this timescale]{pollack96,hubickyj05,lissauer09}.  

\begin{figure*}
\begin{center}
  \includegraphics[width=0.75\textwidth]{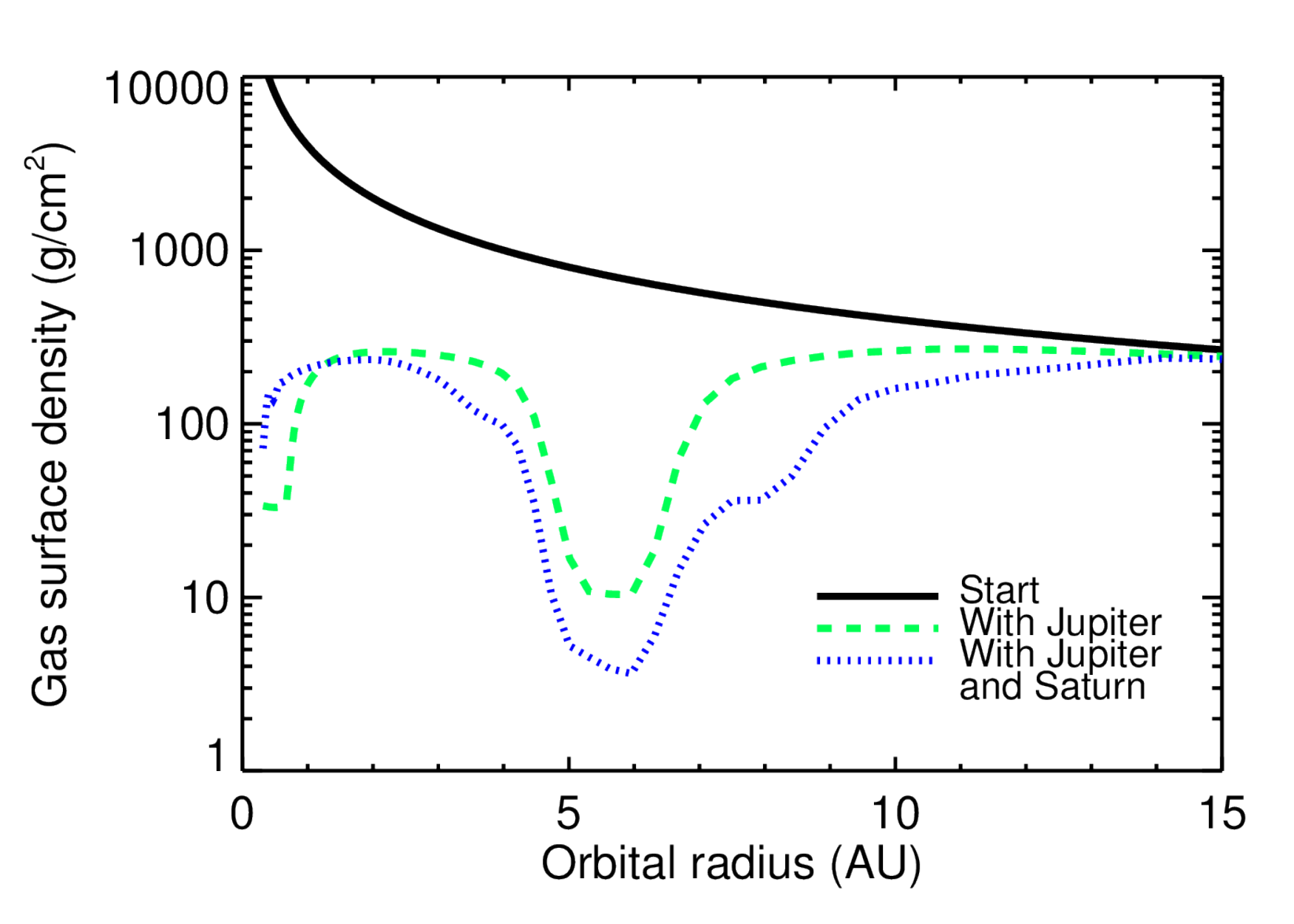}
    \caption[]{Surface density profiles of our underlying gaseous disks, all at their initial values (i.e., not scaled down due to dissipation).  The black curve is the initial disk profile, with surface density $\Sigma \propto r^{-1}$.  The other curves come from hydrodynamical simulations~\citep{morby07a}: the green dashed curve is the disk profile after Jupiter has grown to its full size (at 5.4 AU) and opened a gap and the blue dotted profile is when Jupiter and Saturn are locked in 3:2 mean motion resonance and have opened a common gap.  As our simulations evolve, the disk profile is maintained but the surface density decreases on a prescribed exponential timescale. } 
     \label{fig:gas}
     \end{center}
\end{figure*}

Our initial gas disk profile follows a simple profile, with surface density $\Sigma = 4000 (r/1 AU)^{-1} \, g \, cm^{-2}$ (see Fig.~\ref{fig:gas}). This is a few times more massive than the `minimum-mass solar nebula' model~\citep{weidenschilling77,hayashi81} and consistent with profiles inferred from observations of the outer parts of protoplanetary disks~\citep{andrews09,andrews10}. The disk has a uniform scale height of $H/r = 0.05$. 

We increase the mass of the giant planet cores linearly on a parameterized timescale $\tau_{grow}$ that is generally set to $10^5$~years (although we test the effect of varying $\tau_{grow}$).  This neglects the more complex growth curve envisioned by the core accretion model, in which a core's growth proceeds slowly until runaway gas accretion is triggered, and a slower tail of growth after the planet carves a gap in the disk~\citep{pollack96,ikoma00,rice03,lissauer09}.  Nonetheless, our fiducial timescale of $10^5$ years is close to the Kelvin-Helmholtz timescale for the Jupiter and Saturn region~\citep[e.g.][]{thommes08}.

As the planet grows it modifies the disk's structure by carving an annular gap~\citep{lin86,crida06}. We interpolate the disk's surface density and azimuthal velocity between our initial gas profile and a scaled profile taken from hydrodynamical simulations by \cite{morby07a}, so the gap is carved on the same timescale as the planet's growth.  When Saturn's core grows the process is repeated, interpolating between two surface density profiles from hydrodynamical simulations (Fig.~\ref{fig:gas}).  

The disk dissipates on the exponential timescale $\tau_{gas}$.  During dissipation, the radial surface density profile does not change (apart from the gap opening during the giant planets' growth. Observations of disks in nearby embedded star clusters suggest a disk dissipation timescale of $\sim10^{5-6}$~Myr~\citep{simon95,currie09}. Most of our simulations have a dissipation timescale of $\tau_{gas} = 2-5 \times 10^5$~years, but we also performed simulations in non-dissipating disks to isolate the importance of certain parameters such as the planets' growth timescales.

Planetesimals feel aerodynamic gas drag from the disk, which affects their orbits~\citep{adachi76}. To compute the gas drag acceleration on planetesimals we use the following formula 
\begin{equation}
{\rm {\bf a}_{drag} =- \frac{3C_d \rho_g v_{rel} \bf{v}_{rel}}{4 \rho_p D}}
\end{equation}
where ${\rm C_d}$ is the drag coefficient,  ${\rm \rho_p}$ and ${\rm D}$ are the planetesimal's bulk density (fixed at $1.5 \, g\, cm^{-3}$) and diameter (we tested $D$ from 1 to 1000 km). The  ${\rm v_{rel}}$ vector is the relative velocity of the object with respect to the surrounding gas and ${\rm  \rho_g}$ is the gas density at the planetesimal's immediate location (such that planetesimals on eccentric orbits feel changing gas drag over the course of an orbit). The gas drag coefficient $C_d$ is implemented in a size-dependent fashion~\citep{brasser07}.  

Gravitational interactions with the disk damp the cores' eccentricities and inclinations~\citep{papaloizou00,tanaka04}, although this is only important for core-core interactions (i.e., the simulations with growing Jupiter and Saturn; Sections 4-7).  We start by calculating the damping time scale~\citep{tanaka04}
\begin{equation}
t_{wave} = \frac{1}{\Omega_p}\frac{M_\star}{m_p} \frac{M_\star}{\Sigma_p a^2_p} \left(\frac{H}{r}\right)^4,
\end{equation}
where $\Omega_p$ is the orbital angular velocity, $M_\star$ and $m_p$ are the stellar and planetary mass, respectively, $\Sigma_p$ is the local disk surface density, $a_p$ is the planet's semimajor axis, and $H/r$ the local disk aspect ratio.  

The eccentricity damping timescale $t_e$ is ~\citep{cresswell07}:
\begin{equation}
t_e = \frac{t_{wave}}{0.78}\left[1 - 0.14\left(\frac{e}{H/r}\right)^2 + 0.06 \left(\frac{e}{H/r}\right)^3 + 0.18 \left(\frac{e}{H/r}\right) \left(\frac{i}{H/r}\right)^2\right]
\end{equation}
and the inclination damping timescale $t_i$ is
\begin{equation}
t_i = \frac{t_{wave}}{0.544}\left[1 - 0.30\left(\frac{i}{H/r}\right)^2 + 0.24 \left(\frac{i}{H/r}\right)^3 + 0.14 \left(\frac{e}{H/r}\right)^2 \left(\frac{i}{H/r}\right)\right].
\end{equation}

Our initial conditions include Jupiter and Saturn's cores, each starting at $3 \mearth$, plus 10,000 effectively massless planetesimals spread between 2 and 15-20 AU depending on the simulations, with a 2-3 Hill radius gap around each core.  The cores interact gravitationally with all other bodies. Planetesimals feel the cores' gravity but do not self-gravitate.  Simulations were integrated for 0.5-3 Myr with a timestep of 0.1 years.  Particles were removed from the simulation if they were closer than 0.35 AU from the Sun or farther than 200 AU.

We tested the effect of the following parameters: the planetesimal size, the gas accretion timescale, the disk lifetime, and Jupiter and Saturn's growth and migration histories.

\section{In-situ growth of Jupiter and Saturn}
\begin{figure}
\includegraphics[width=0.95\textwidth]{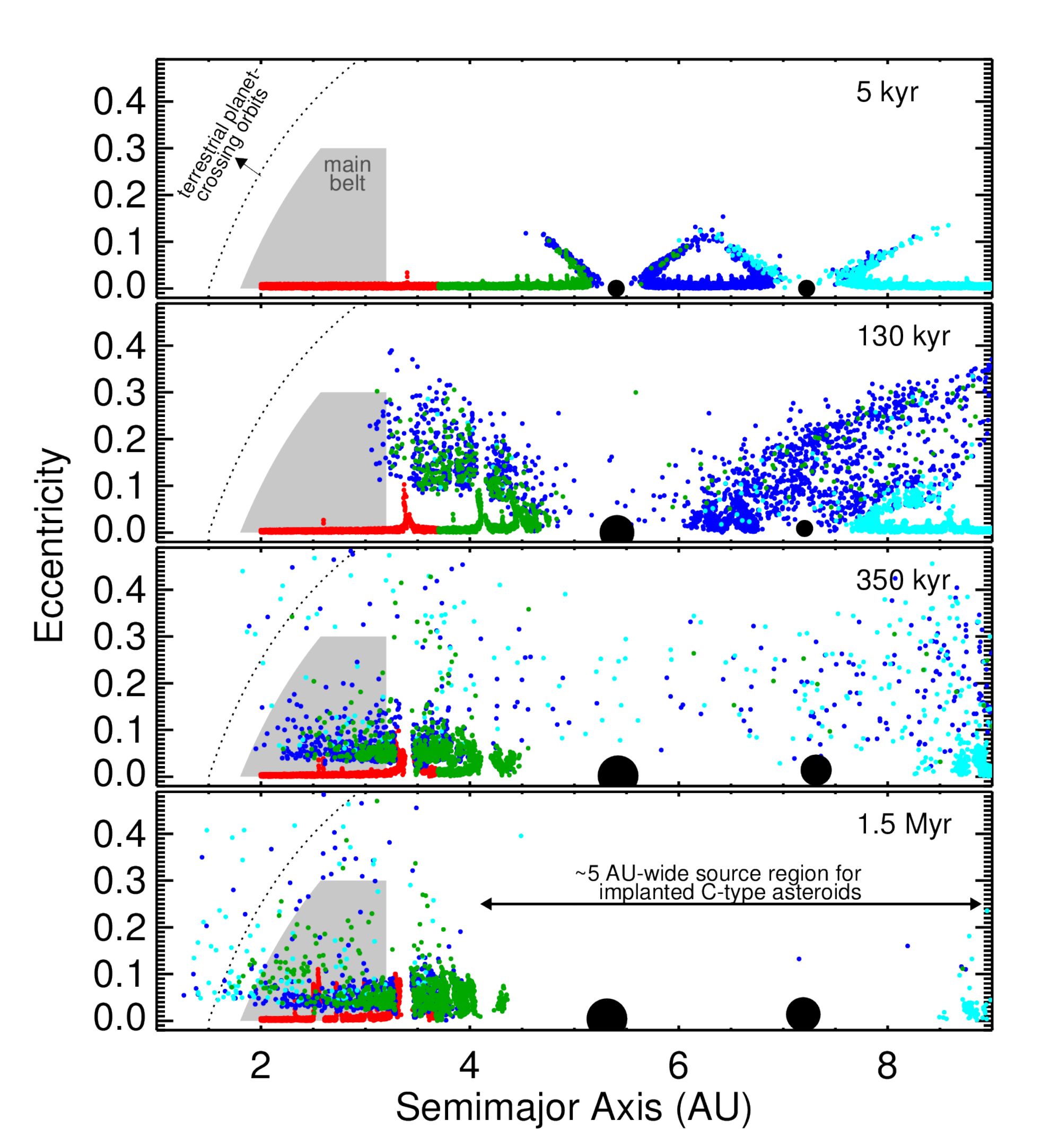}
    \caption[]{Snapshots of a simulation of Jupiter and Saturn's in-situ gas accretion. Planetesimals are color-coded simply by their starting orbital radius. Here Jupiter's grew linearly from a $3 \mearth$ core to its full mass from 100-200 kyr and Saturn from 300-400 kyr.  The planets were locked in 3:2 mean motion resonance throughout the simulation, a natural outcome of convergent migration~\citep{masset01,pierens08,pierens14}. In this case, planetesimals were 100~km in diameter, the gas disk depleted uniformly in radius on an exponential time scale $\tau_{gas} = 2\times10^{5}$~years. The main asteroid belt is shaded in gray. Above and to the left of the dotted curve, planetesimals have perihelion distances that cross the terrestrial planet region. } 
     \label{fig:jupevol}
\end{figure}

Figure~\ref{fig:jupevol} shows a simulation with Jupiter and Saturn growing in a sea of 100~km planetesimals. At early times cores scatter nearby planetesimals, which follow ``wings" in semimajor axis-eccentricity space corresponding to regions of constant Tisserand parameter~\citep{levison97}. Given the cores' small masses they only excite eccentricities up to $\sim 0.1$. During Jupiter's rapid gas accretion (starting at $10^5$ years) the mass-dependent orbital stability limit~\citep{marchal82,gladman93} shifts. Nearby planetesimals' orbits are destabilized and gravitationally scattered by Jupiter, again along Tisserand wings but to high eccentricity orbits that crossed the Solar System. The large velocity differential with respect to the gas disk generates a drag force that re-circularizes planetesimals' orbits~\citep{adachi76} and deposits many into the main asteroid belt.  Planetesimal scattering was repeated when Saturn's core underwent rapid gas accretion from 300-400 kyr in Fig.~\ref{fig:jupevol}. By this point the gas disk was depleted and gas drag was weaker. Many planetesimals were scattered by the growing Saturn and subsequently so strongly by Jupiter that they overshot the asteroid belt and crossed the growing terrestrial planets' orbits.  

\subsection{Effect of planetesimal size}

We use four simulations to illustrate more details of the asteroid implantation process.  These simulations are identical to the simulation presented in Fig.~\ref{fig:jupevol} but for four different planetesimal sizes: diameters $D=$1, 10, 100, and 1000~km. Each simulation was run for 3 Myr with an exponential gas dissipation timescale of 200 kyr.  Jupiter grew from 100-200 kyr and Saturn from 300-400 kyr.

\begin{figure*}
\includegraphics[width=0.49\textwidth]{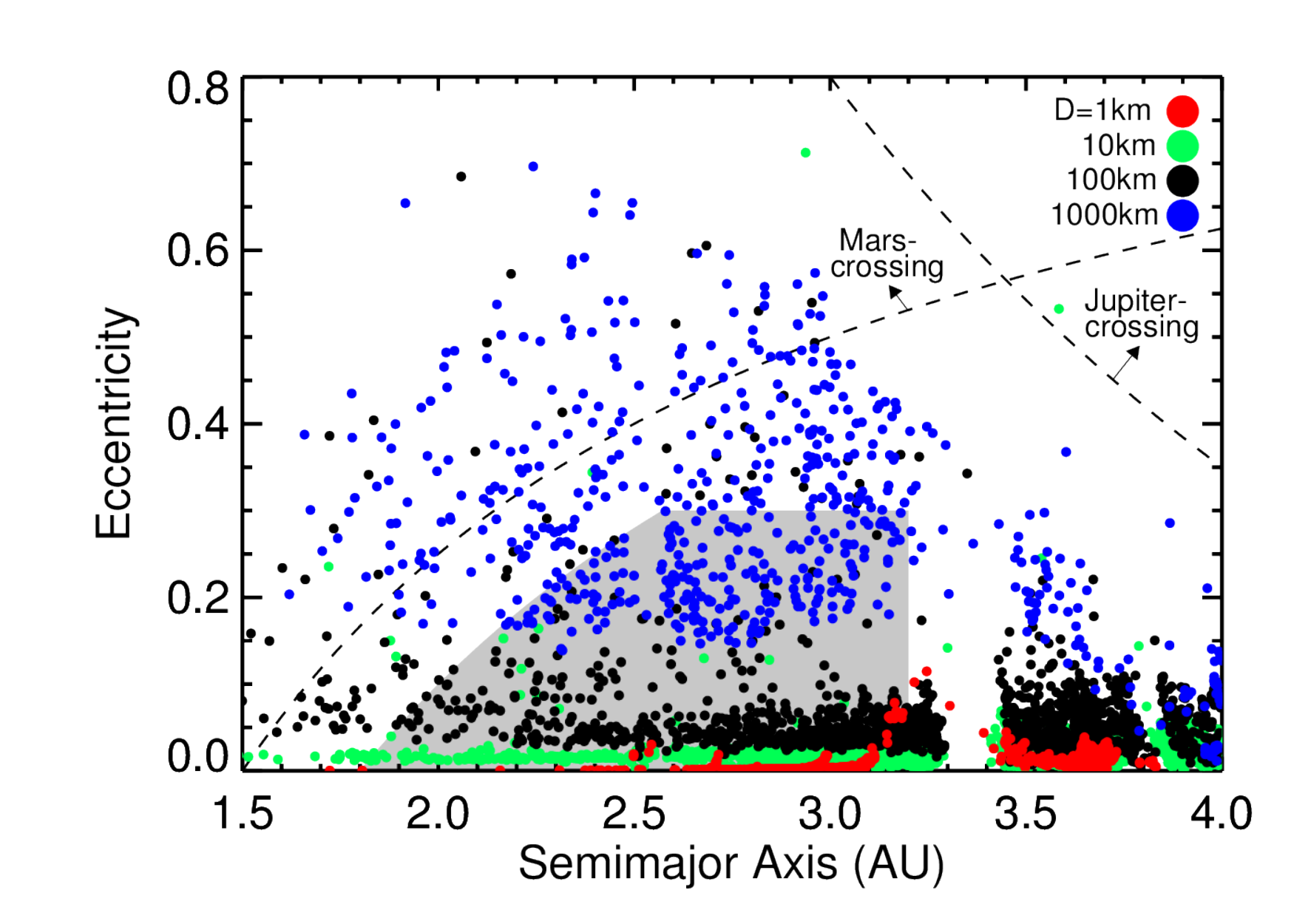}
\includegraphics[width=0.49\textwidth]{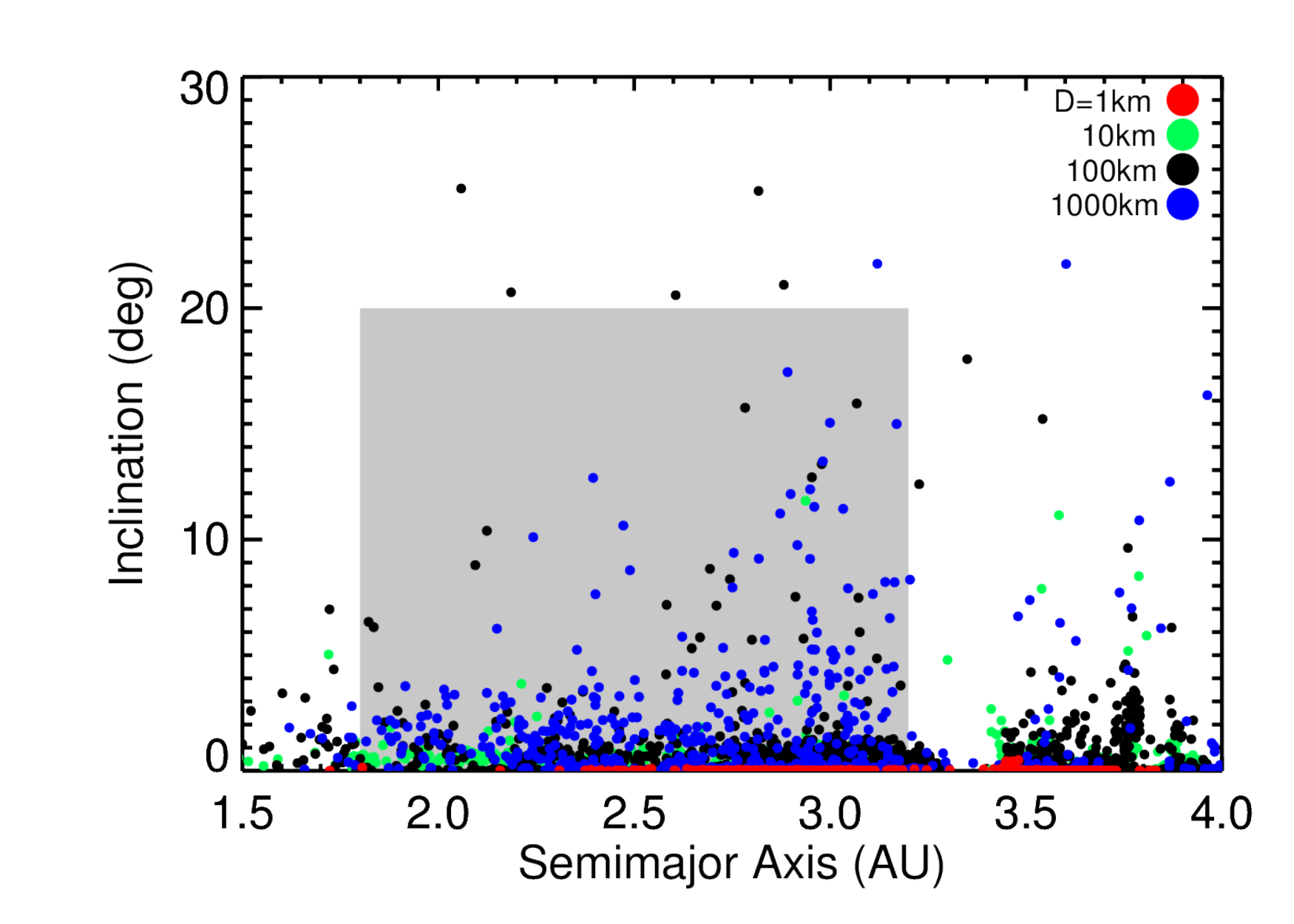}
   \caption[]{Orbital distribution of planetesimals scattered into the inner Solar System from four simulations with different planetesimal sizes.  The main asteroid belt -- defined here to have perihelion $q > 1.8$~AU, eccentricity $e<0.3$ and semimajor axis $a < 3.2$~AU -- is shaded. The dashed lines show orbits that cross that of Jupiter and Mars (note that Mars is not included in these simulations).   }
     \label{fig:long_2}
\end{figure*}

Figure~\ref{fig:long_2} shows the orbital distribution of planetesimals scattered into the inner Solar System in the four simulations. There is a segregation in orbital eccentricity by planetesimal size.  Large planetesimals feel much weaker gas drag and require a longer time for their eccentricities to damp. Scattered planetesimals are generally captured into the main belt by having their eccentricities damped at near-constant semimajor axis. Larger planetesimals take longer for their eccentricities to be damped and so are generally deposited on higher-eccentricity (and inclination) orbits. In the simulations from Fig.~\ref{fig:long_2}, all but the largest planetesimals have damped to near-circular orbits.  As their eccentricities damp slowly, larger planetesimals tend to undergo more close encounters with the Jupiter (compared with smaller planetesimals) and are more readily scattered onto very high-eccentricity, close-in orbits that cross the inner Solar System.

The eccentricities and inclinations of all but the largest implanted planetesimals in Fig.~\ref{fig:long_2} are significantly lower than in the present-day belt. A later dynamical phase is required to excite the asteroids' orbits to the current observed levels. This may or may not be linked with a depletion of the belt. This is discussed in Section 8.3.

\begin{figure}
\begin{center}
\includegraphics[width=0.75\textwidth]{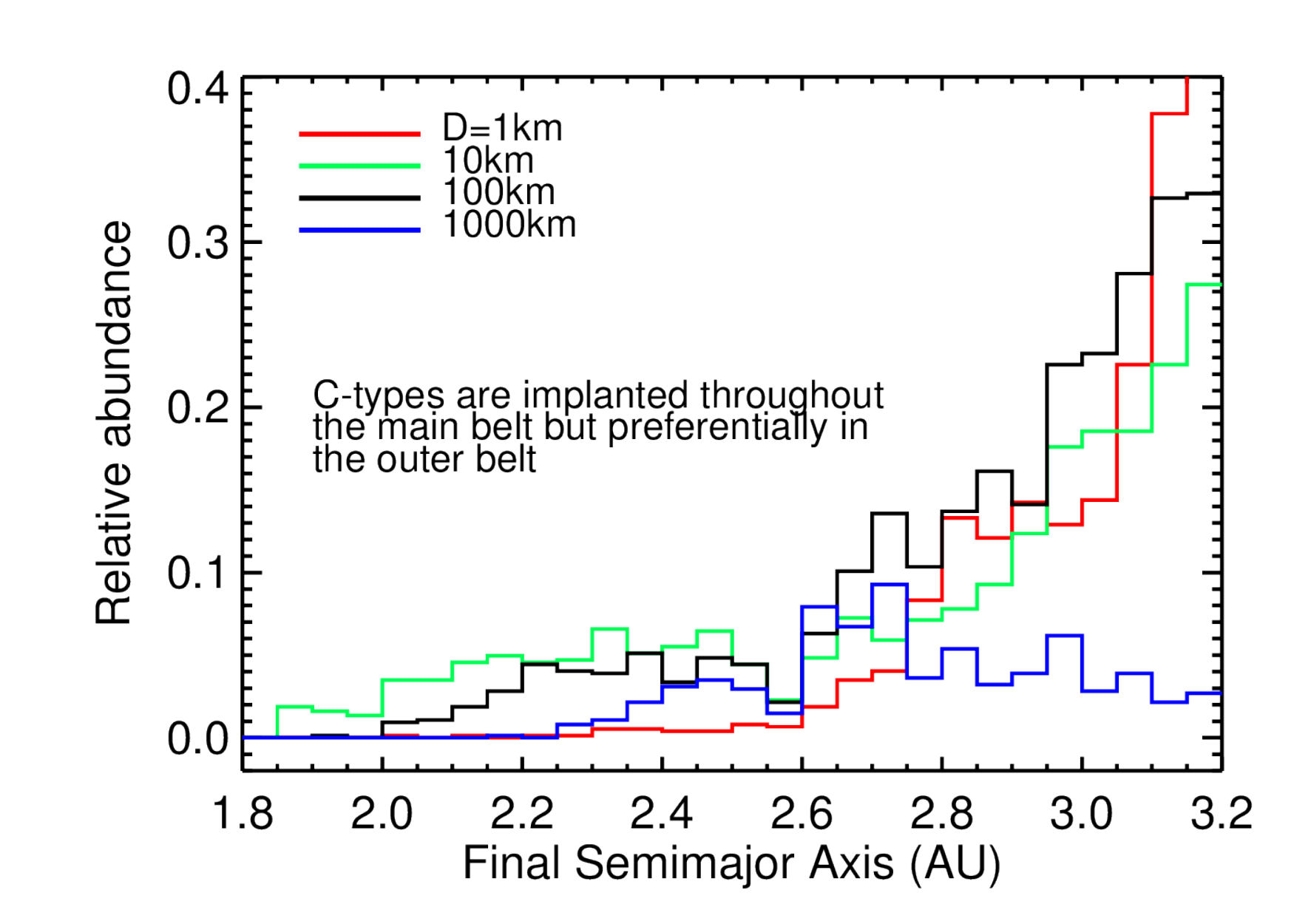}
    \caption[]{Radial distribution of planetesimals after they were implanted from beyond 4 AU into the main asteroid belt from a set of twelve simulations, four each with planetesimal diameters $D$ of 1, 10, 100, and 1000~km.  Each color corresponds to a given planetesimal size. The distribution for each population was normalized a common value such that the distributions indicate the relative abundance of different sizes in each location. The normalization was for the $D=1$~km planetesimals in the outermost bin. } 
     \label{fig:jsdistr}
     \end{center}
\end{figure}

With the simple assumption that planetesimals originating past 4 AU have C-type compositions, and that S-types represent planetesimals native to the main belt, our simulations broadly match the observed S/C-type dichotomy. Scattering promotes the implantation of medium-sized (10~km and 100~km) planetesimals throughout the main belt, with a distribution that rises toward the outer main belt (Fig.~\ref{fig:jsdistr}). The smallest planetesimals feel the strongest gas drag and quickly decouple from Jupiter after they are scattered, implanting them closer to Jupiter than larger planetesimals. The distribution of the largest (1000~km) planetesimals has a broad peak centered at $\sim 2.7$~AU. As gas drag is weaker, large planetesimals' eccentricities damp slowly so they undergo many scattering events with Jupiter.  Planetesimals beyond $\sim 2.7$~AU can re-encounter Jupiter (since their aphelia $Q = a (1+e) = 5.4$~AU for $e=1$) and are cleared out by scattering whereas those at smaller orbital radius are protected. The confluence of these factors produces a peak at half of Jupiter's orbital radius. Ceres, the only $\sim$1000~km asteroid, lies close to this peak (at 2.77 AU).

\begin{figure*}
\includegraphics[width=0.49\textwidth]{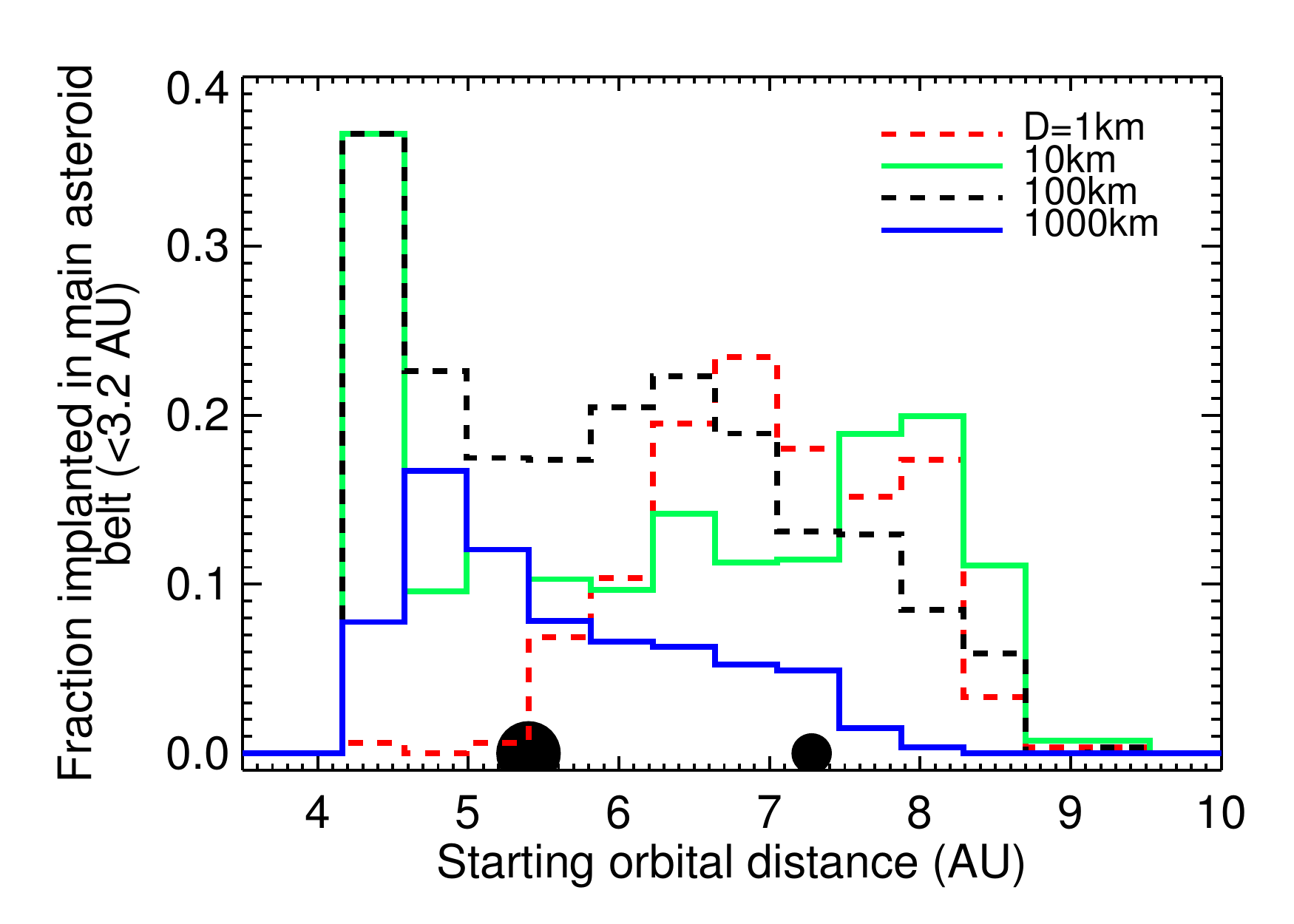}
\includegraphics[width=0.49\textwidth]{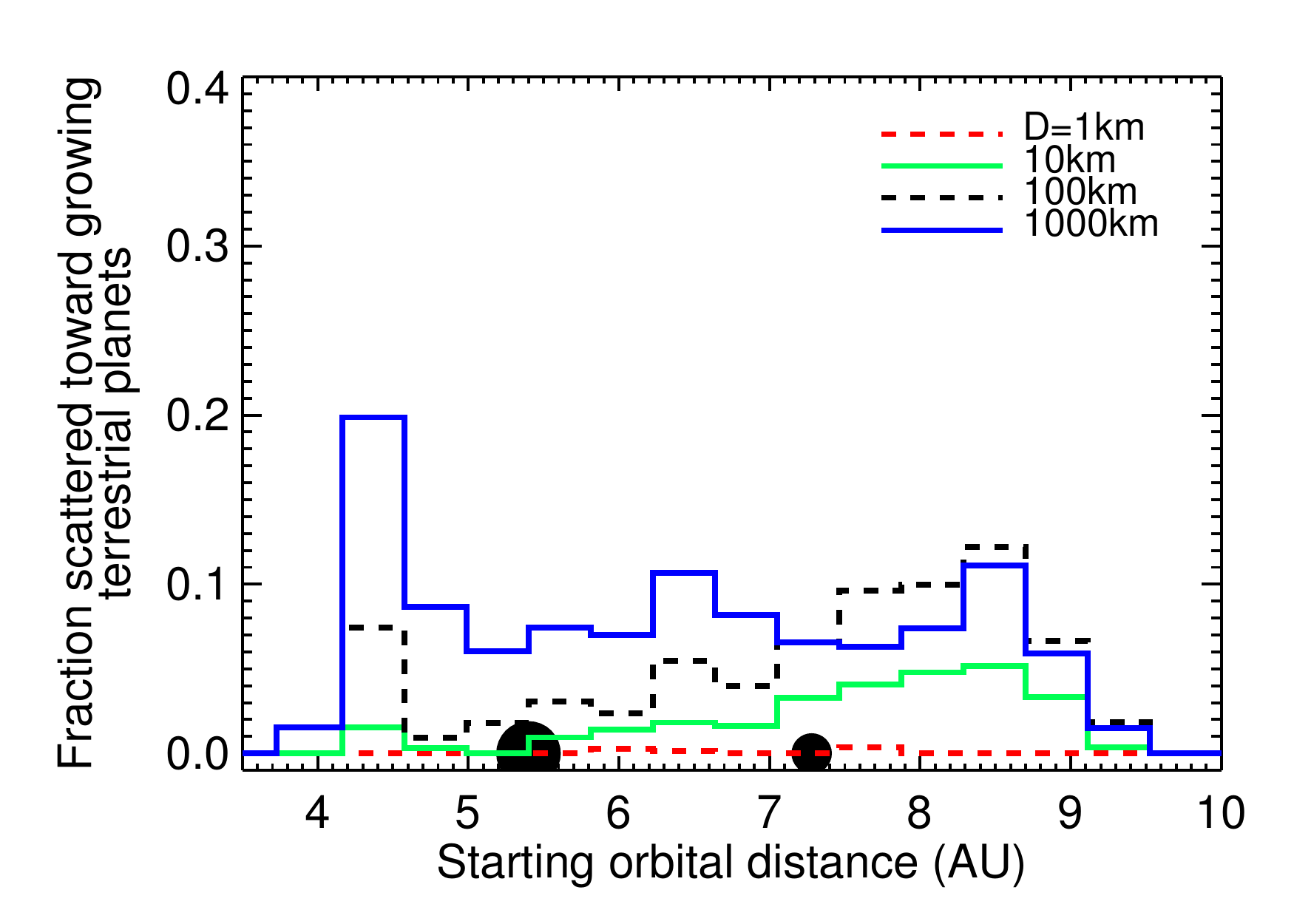}
\caption[]{{\bf Left:} The efficiency with which planetesimals were implanted onto stable orbits within the main belt as a function of the planetesimals' initial orbital radius.  In other words, the source region of C-types in the simulations with Jupiter and Saturn growing in-situ. {\bf Right:} Source region of potential water-delivering planetesimals.  These are objects that were scattered onto orbits crossing the orbit of present-day Mars, i.e., to a minimum perihelion distance of 1.5 AU or smaller.  The format and scale is the same in the two panels.  }
     \label{fig:imp_scat}
\end{figure*}

From the same four simulations, Fig.~\ref{fig:imp_scat} shows the source regions for planetesimals that were implanted into the main asteroid belt (left) as well for planetesimals that were scattered {\it past} the asteroid belt onto orbits crossing those of the growing terrestrial planets.  A comparison between the two panels shows that: 
\begin{enumerate}
\item Overall, more planetesimals were implanted into the asteroid belt than scattered to the terrestrial planets in these simulations.  This is simply because the giant planets' growth happened relatively early in the disk's lifetime, favoring implantation over water delivery (see Section 4).  A later phase of planetesimal scattering driven by, for example, the migrating ice giants~\citep{izidoro15b}, could easily serve to deliver more water to the terrestrial planets or pollute the belt.
\item Moderate (1-100~km) sized planetesimals are most easily implanted into the asteroid belt whereas the largest (1000~km) planetesimals are more readily scattered into the inner Solar System. This is explained by the strong size-dependence of aerodynamic gas drag. The eccentricities of the largest planetesimals are damped the slowest, so that these large planetesimals undergo a higher number of scattering events with Jupiter and reach higher eccentricities (and correspondingly lower perihelion distances) than smaller planetesimals that feel stronger gas drag.
\item The radial distribution of implanted and water-delivering planetesimals tells a story of timing.  For $D \le 100$~km, asteroidal implantation occurs when both Jupiter and Saturn grow. For the largest planetesimals, implantation is most effective when Jupiter grows, when the gas density is highest.  

The $D=1000$~km planetesimals are scattered to the terrestrial region throughout the simulation such that the source region of water-delivering bodies is roughly flat with initial orbital radius.  Smaller ($D \le 100$~km) planetesimals are only scattered to such high eccentricities and small semimajor axes late in the simulation when the gas density was lower. They therefore have an increase in the efficiency of water delivery near Saturn's orbital radius.  In fact, near Saturn's orbital radius the efficiency of water delivery is comparable for 10, 100 and 1000~km planetesimals.  
\end{enumerate}

\subsection{Effect of the giant planets' growth timescale and the gas disk's dispersal timescale}
We performed additional simulations to test the effect of the giant planets' formation timescale and the disk's dispersal timescale.  

To isolate the effects of the planets' growth timescale, we first ran simulations of a single Jupiter growing in a non-evolving disk.  As the planet grew it carved a gap in the disk (see Fig.~\ref{fig:gas}) but the disk as a whole did not dissipate.  We tested three different timescales for (linear) growth: $10^4$, $10^5$, and $10^6$ years.  We ran identical simulations with planetesimals of different sizes for each growth timescale.  Each simulation was run for 1.2 Myr, with the giant planet's growth starting at 0.1 Myr.  By the end of each simulation the vast majority ($>95\%$) of planetesimals had been removed from the region between 4 and 7 AU, and there was little evolution during the last $10^5$ years.

\begin{figure*}
\begin{center}
\includegraphics[width=0.75\textwidth]{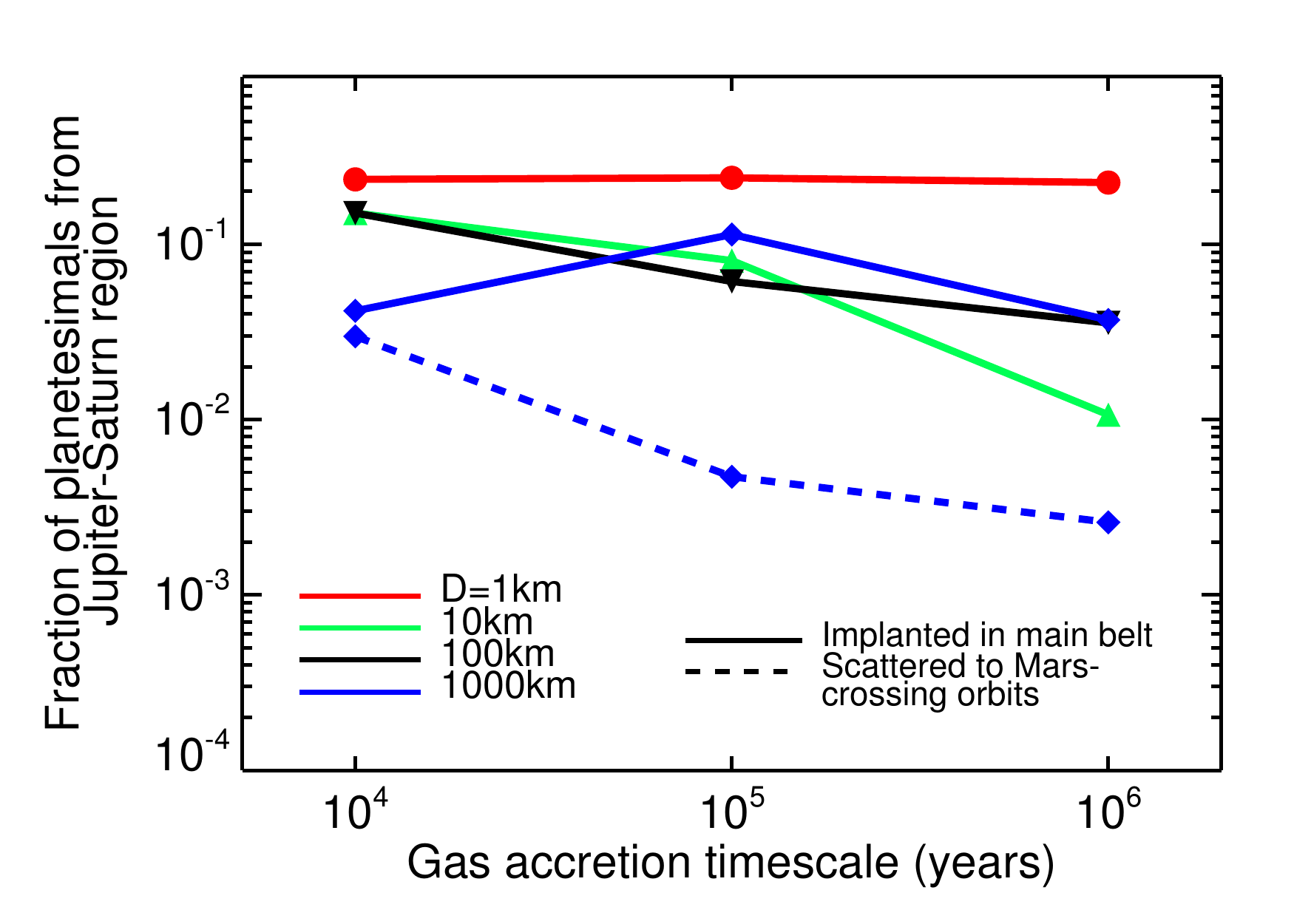}
    \caption[]{Implantation and scattering efficiency in simulations with only Jupiter growing in-situ in a non-dissipating disk.  The solid lines show the implantation efficiency for different-sized planetesimals.  The blue dashed line shows the efficiency of scattering onto Mars-crossing orbits.  Given the high-density, non-evolving disk, only the largest planetesimals were scattered toward the terrestrial region in these simulations.}
\label{fig:tgrow}
\end{center}
\end{figure*}

Figure~\ref{fig:tgrow} shows that Jupiter's growth timescale has only a modest effect on the implantation process.  The efficiency of implantation into the main belt did not vary by more than a factor of a few for the smallest and largest planetesimals, but it differed by a factor of 5-10 for between the shortest and longest growth timescales for medium-sized ($D=$10-100~km) planetesimals.  This is because with a slow-growing gas giant scattered planetesimals have longer to interact with the gas disk and so they are implanted closer to the giant planet, exterior to the main belt.  For the same reason, the fraction of large planetesimals scattered onto Mars-crossing orbits also dropped for longer growth timescales. The small planetesimals are basically unaffected because they preferentially drift into the main belt rather than scatter.  

For the same growth timescale of Jupiter ($10^5$~years), the efficiency of asteroidal implantation in a static disk (Fig.~\ref{fig:tgrow}) is within a factor of 2-3 of that in a dissipating disk (Fig.~\ref{fig:imp_scat}). We conclude that the disk's dissipation timescale is not a key parameter in the implantation mechanism, at least for the combination of parameters that we have tested. 

\subsection{Implantation of planetesimals into the outer Solar System}

During the gas giants' growth, some planetesimals are scattered outward.  Under the action of eccentricity damping from gas drag, a fraction are captured onto stable orbits exterior to Saturn's. 

\begin{figure*}
\includegraphics[width=0.49\textwidth]{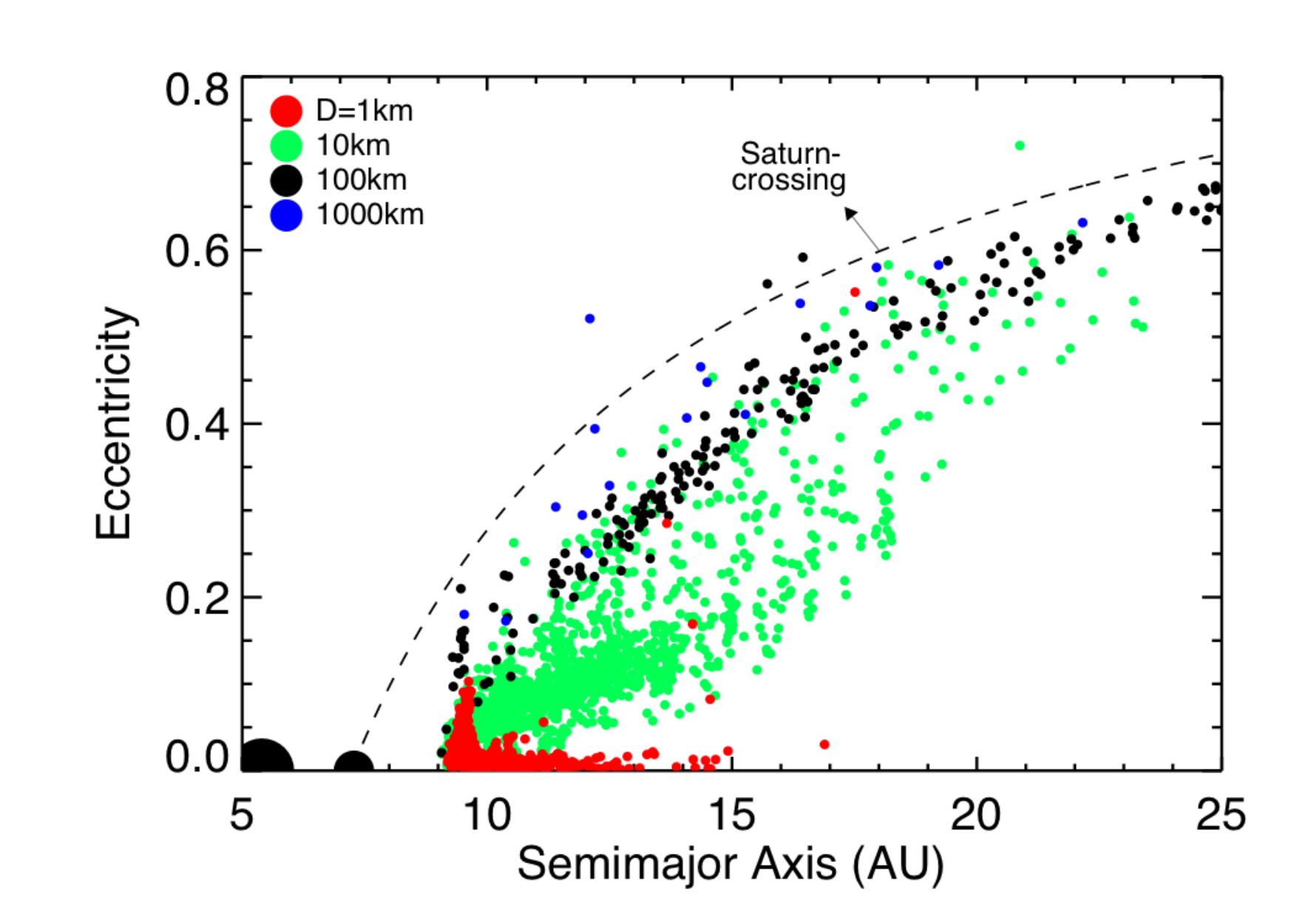}
\includegraphics[width=0.49\textwidth]{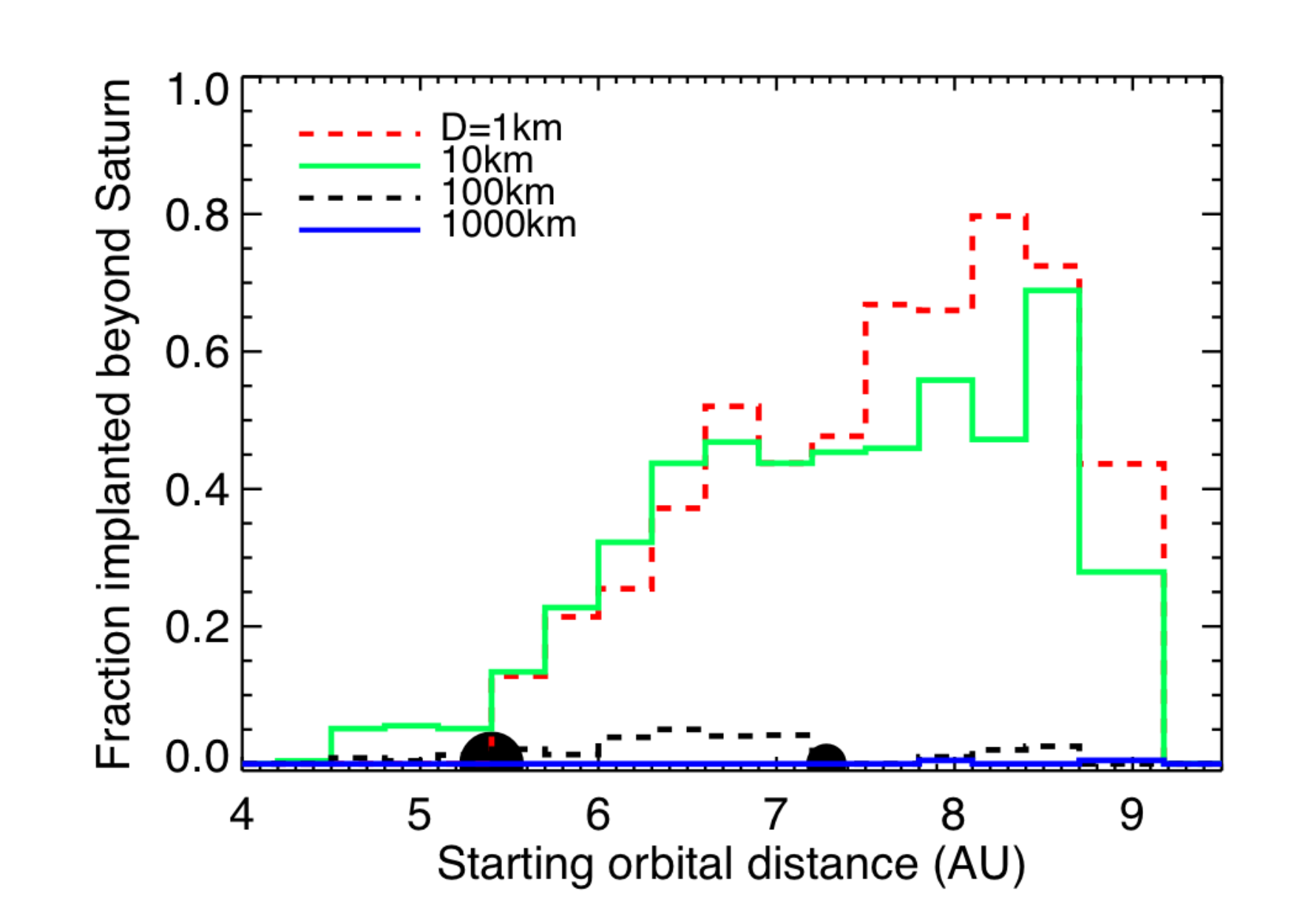}
\caption[]{{\bf Left:} Orbital distribution of planetesimals scattered past Saturn (including unstable ones with orbits that cross Saturn's) from the same simulations as in Figs.~\ref{fig:jsdistr} and~\ref{fig:imp_scat}. {\bf Right:} Source region of planetesimals implanted onto stable orbits beyond Saturn, defined as having perihelia larger than 9 AU.  }
     \label{fig:scatout}
\end{figure*}

Figure~\ref{fig:scatout} shows the planetesimals scattered outward in the same simulations from Figs~\ref{fig:jsdistr} and~\ref{fig:imp_scat}.  As for the inner Solar System, planetesimals are scattered along curves of constant Tisserand parameter extending to arbitrarily high eccentricity. Gas drag acts to decrease their eccentricities and strand them on stable orbits.  Given the lower density of the outer parts of the disk, this process is more strongly size-dependent than implantation into the main belt.  While 1 and 10~km planetesimals were scattered outward at a rate of $\sim$50\% from Saturn's neighborhood, the process was far less efficient for large planetesimals.  The reason is simple: for large planetesimals the timescale for eccentricity damping due to gas drag is longer than the timescale for ejection from the Solar System.  As a result, large planetesimals are preferentially lost whereas small ones are captured.  This may have consequences for the expected size distribution of the Oort cloud~\citep{brasser07}. 

Smaller (1 and 10~km) planetesimals are preferentially implanted closer to Saturn, with semimajor axes between 10 and 15 AU.  A fraction are scattered to more distant orbits but with large eccentricities.  These objects certainly contaminate the outer Solar System with material from the Jupiter-Saturn region. Although these simulations do not include the ice giants (or their precursors), we expect them to participate in the same process. A fraction of planetesimals scattered outward onto ice giant-crossing orbits should be scattered outward, and the process should continue until a fraction of planetesimals reaches past the outermost planet (or perhaps a wide gap between planets if one exists). See Section 7.

If there is a strong gradient in composition across the Jupiter-Saturn region then it is possible that this is a dynamical mechanism to deposit refractory objects into the outer Solar System, either by direct implantation or by the disruption of weak planetesimals on eccentric orbits that pollute the outer Solar System with a trail of debris.  However, given the efficiency of implantation into the outer main belt we expect most of this material to have carbonaceous composition (like the C-types) so it is unclear whether this mechanism can explain the origin of refractory grains in comets such as those seen by the Stardust mission~\citep{brownlee06} or in the Oort cloud~\citep{meech16}.  

\section{Timing of giant planet formation: water delivery vs. asteroidal implantation}

We now test the importance of the timing of giant planet formation. Timing matters because the gas disk's density decays rapidly, and most planet-forming disks are gone within a few million years~\citep{haisch01,mamajek09}. 

We performed simulations in which Jupiter and Saturn grew simultaneously in place (without migrating) on a $10^5$ year timescale in a disk that itself dissipated on an exponential timescale of $2.5 \times10^5$ years. We varied the start of the giant planet's gas accretion from zero to 1 Myr, i.e., from 0 to 4 e-folding timescales in the disk's density. Simulations contained $10^4$ 100~km planetesimals initially between 4 and 9 AU and were integrated for 2.5 Myr (10 e-folding times of the gas' surface density). 

\begin{figure}
\begin{center}
\includegraphics[width=0.75\textwidth]{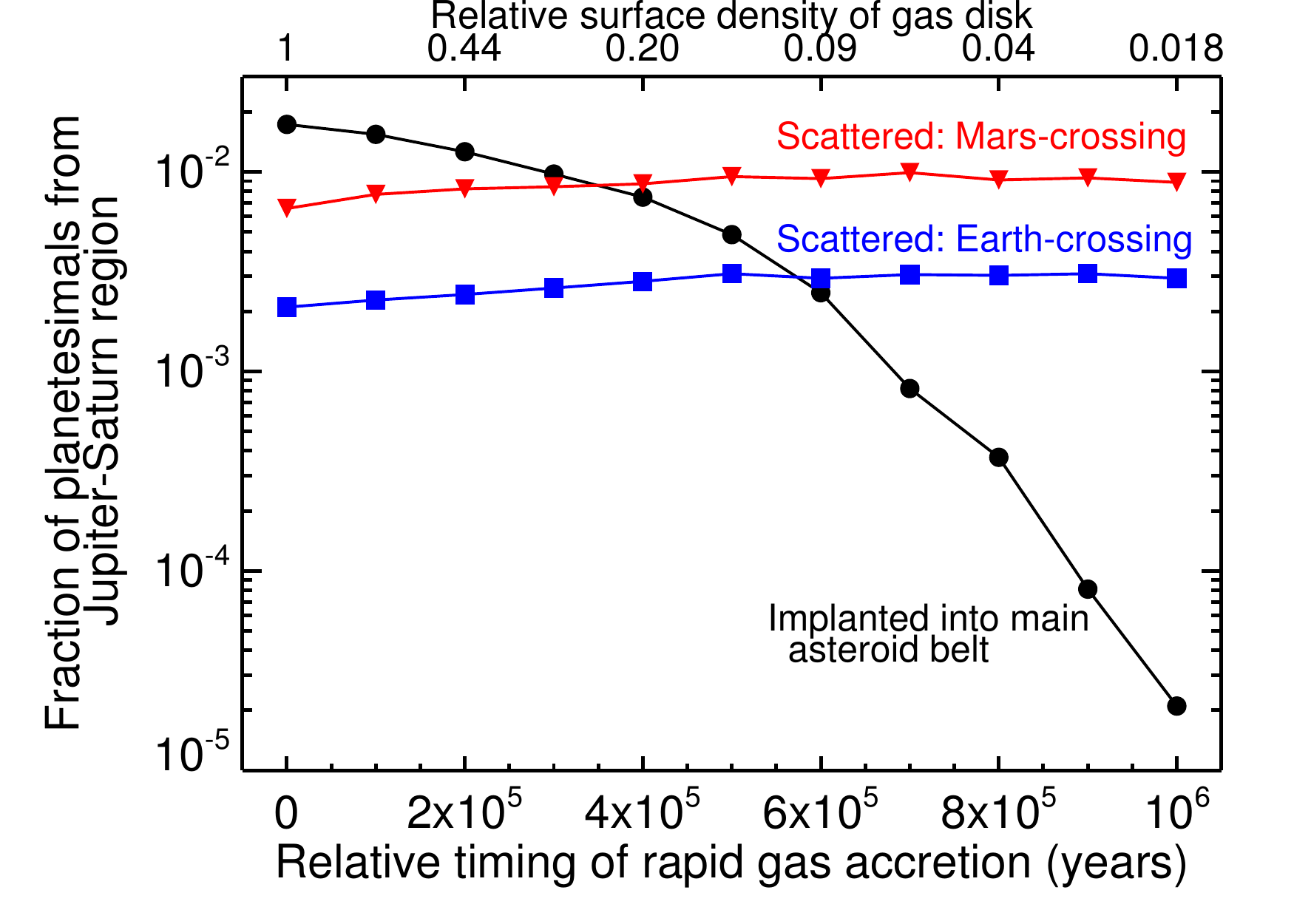}
    \caption[]{The relative importance of C-type asteroid implantation and water delivery as a function of the timing of giant planet formation. The y axis represents the fraction of scattered planetesimals that are implanted into the main asteroid belt (black curve) or past the asteroid belt toward the terrestrial planet zone (red curve indicates a minimum perihelion of 1.5 AU and blue curve of 1 AU). The disk's exponential dissipation timescale was $2.5 \times10^5$ years, so the timing of gas accretion is linked with the disk's surface density (see top x axis). Jupiter and Saturn grew in-situ on a 100 kyr timescale, and all planetesimals were 100~km in diameter.  } 
     \label{fig:water}
\end{center}
\end{figure}

Figure~\ref{fig:water} shows that the balance between implantation into the asteroid belt and scattering to the terrestrial zone is a strong function of the timing of giant planet formation (Fig.~\ref{fig:water}).  At early times, scattered planetesimals' eccentricities are quickly damped, such that implantation is significantly more efficient than scattering into the inner Solar System.
 
The efficiency of asteroid belt implantation drops strongly as the disk dissipates (Fig.~\ref{fig:water}). Scattering to the terrestrial zone becomes more prevalent for simulations in which the giant planets formed later than $\sim 5 \times 10^5$~years ($= 2 \tau_{gas}$). Scattered planetesimals preferentially reach high-eccentricity orbits late in the disk lifetime. 

Because gas drag is size-dependent~\citep{adachi76}, the relative contribution is tipped from asteroidal implantation to terrestrial planet-crossing for different size planetesimals at different times. Due to their weaker gas drag, large planetesimals are strongly scattered toward the terrestrial region at early times whereas small planetesimals are only scattered strongly at late times when the disk is low-density.  

Figure~\ref{fig:water} illustrates a specific setup but can be interpreted in a more general sense.  The timing of rapid gas accretion can be thought of as simply the time at which a planetesimal is scattered.  In this case the planetesimals originated in the Jupiter-Saturn region, but other sources can be imagined.  For example, the growth and migration of the ice giants~\citep{izidoro15c} would have scattered planetesimals toward the giant planets (see Section 7).  The fate of scattered planetesimals -- and whether they likely contributed to the C-types in the main belt or Earth's water budget -- can be broadly determined by knowing the timing of the scattering.

\section{Different scenarios for Jupiter and Saturn's migration}

Orbital migration of planets embedded in gaseous protoplanetary disks is unavoidable~\citep{goldreich80,lin86,ward97,armitage07,paardekooper11}.  Yet Jupiter and Saturn's migration history is only loosely constrained. Given the rapid migration of low-mass planets, it is likely that Jupiter's core underwent large-scale migration. Combining pebble accretion and migration, \cite{bitsch15} showed that Jupiter's core may have originated as far as 20-30 AU from the Sun and migrated inward to be stranded near 5 AU. In contrast, \cite{raymond16} proposed that Jupiter's core may have formed very close to the Sun and migrated outward.  

This panel of possible migration histories gets even wider when Saturn is accounted for.  The Grand Tack model proposes one plausible migration history for the giant planets, in which Jupiter's inward migration was reversed by Saturn's growth and capture in 2:3 mean motion resonance~\citep{walsh11}. This mechanism has been validated by a number of hydrodynamical studies~\citep{masset01,morby07b,pierens08,crida09,zhang10,pierens11}. By calculating a self-consistent disk structure, \cite{pierens14} showed that Jupiter and Saturn have a wider range of evolutionary pathways. For higher disk masses and higher viscosities, Jupiter and Saturn enter 3:2 resonance and migrate outward, as in previous studies in isothermal disks.  However, for modestly lower disk masses and viscosities Saturn's migration is slower and it is trapped in 1:2 rather than 2:3 resonance with Jupiter.  In this configuration the two planets may either migrate outward or remain on roughly stationary orbits~\citep{pierens14}.  Here we focus on inward migration (or roughly stationary outcomes for the in-situ growth case from Section 3) and discuss the effect on the Grand Tack model below.

\begin{figure}
\begin{center}
\includegraphics[width=0.49\textwidth]{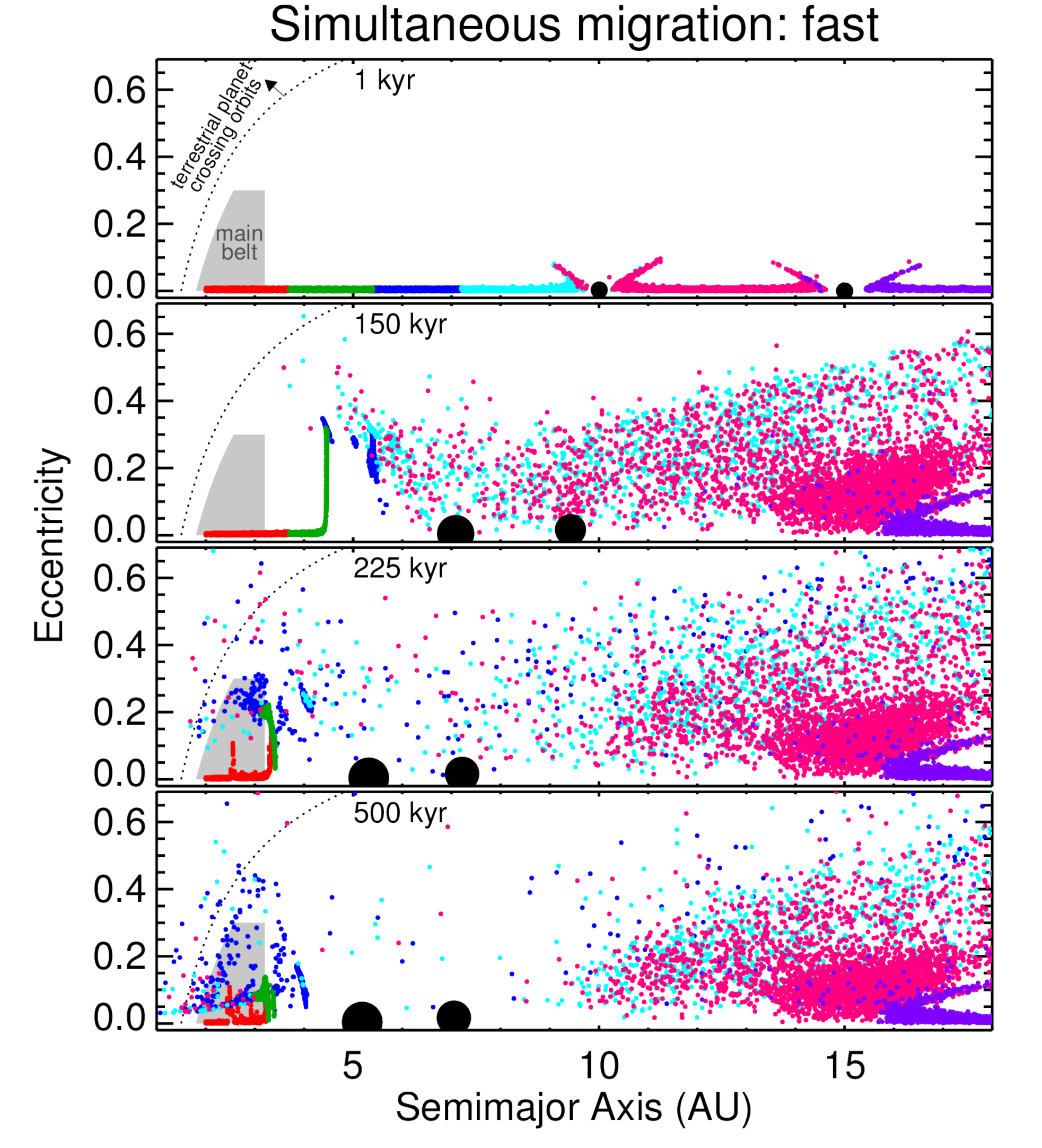}
\includegraphics[width=0.49\textwidth]{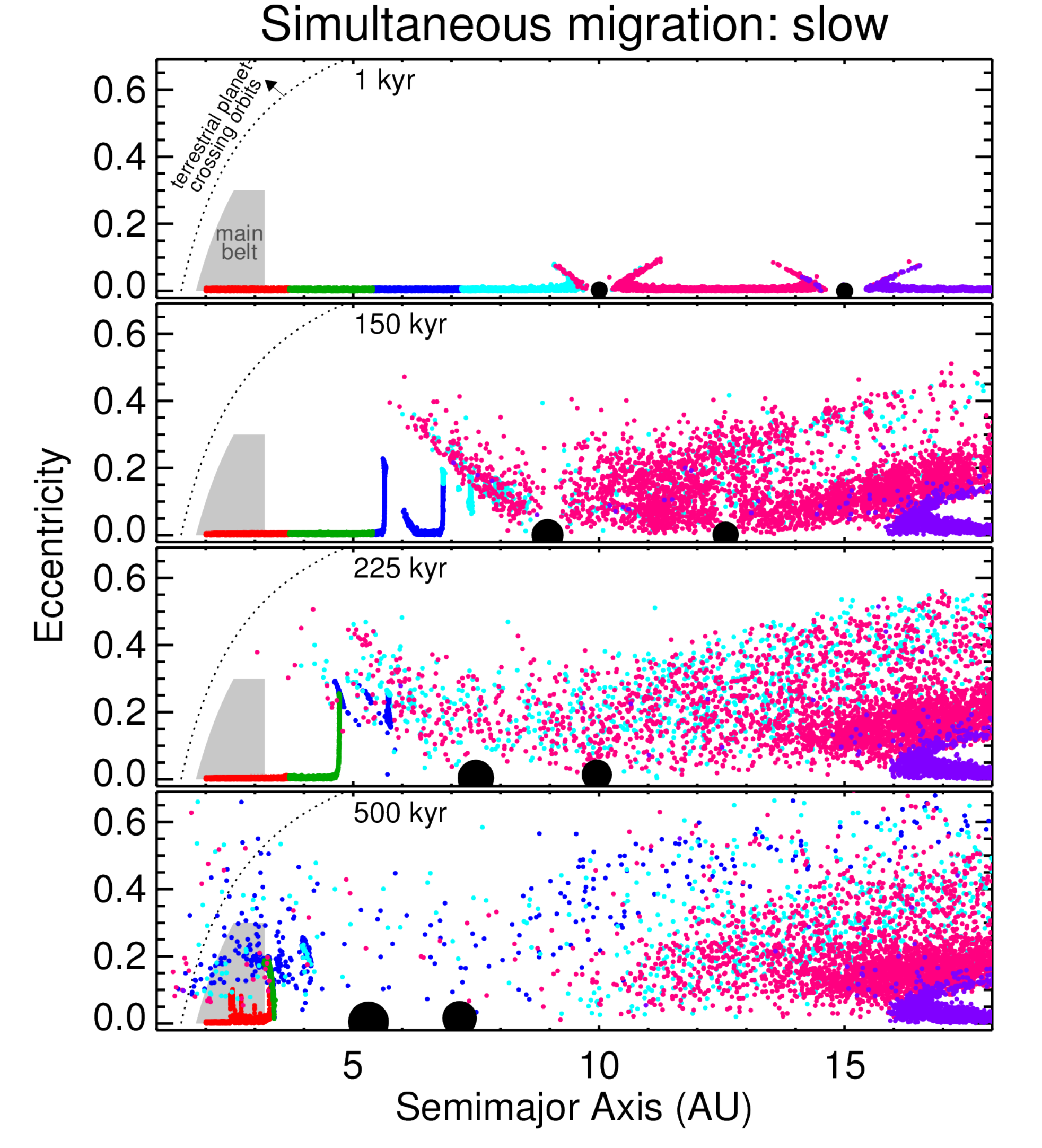}
\includegraphics[width=0.49\textwidth]{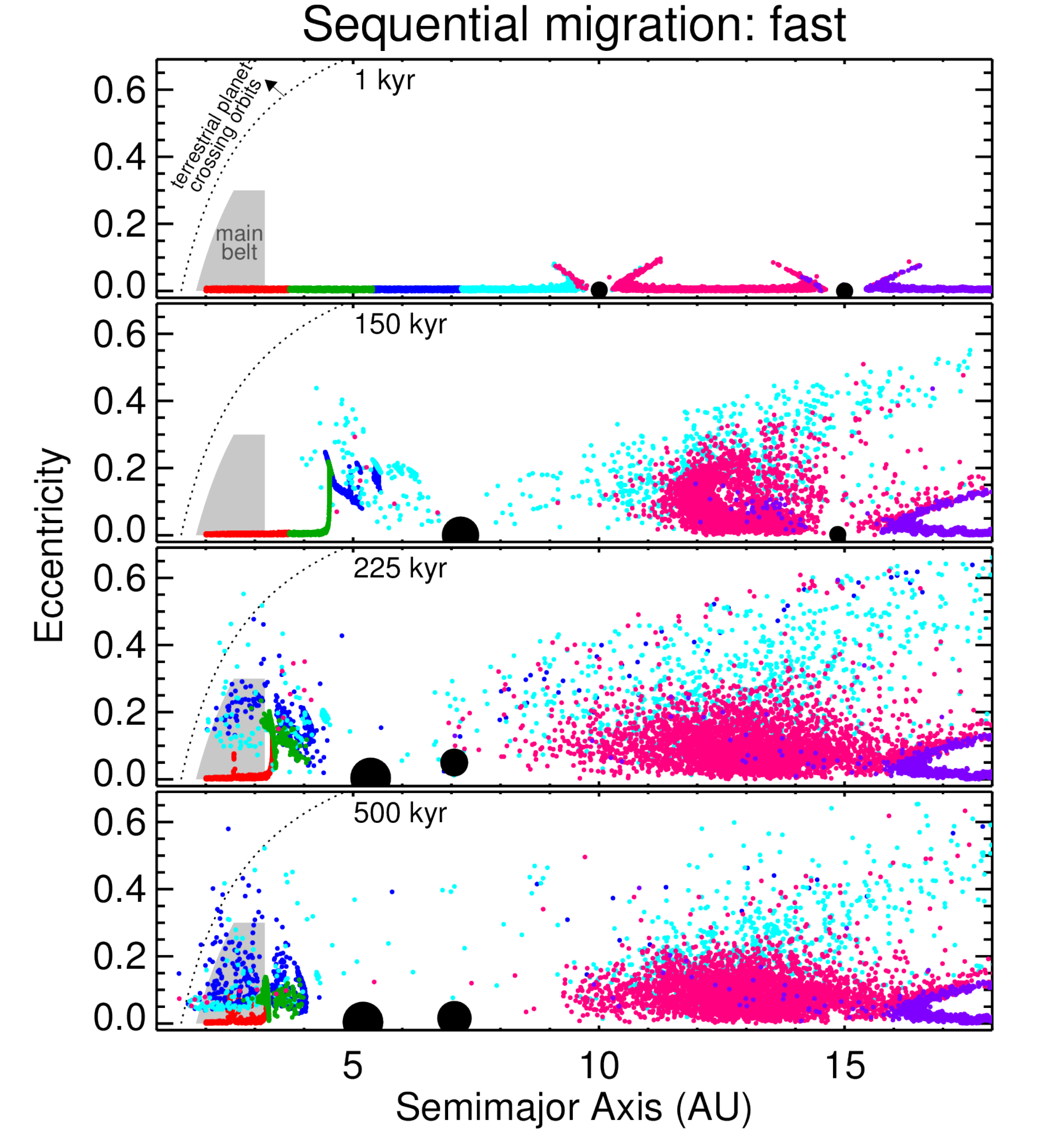}
\includegraphics[width=0.49\textwidth]{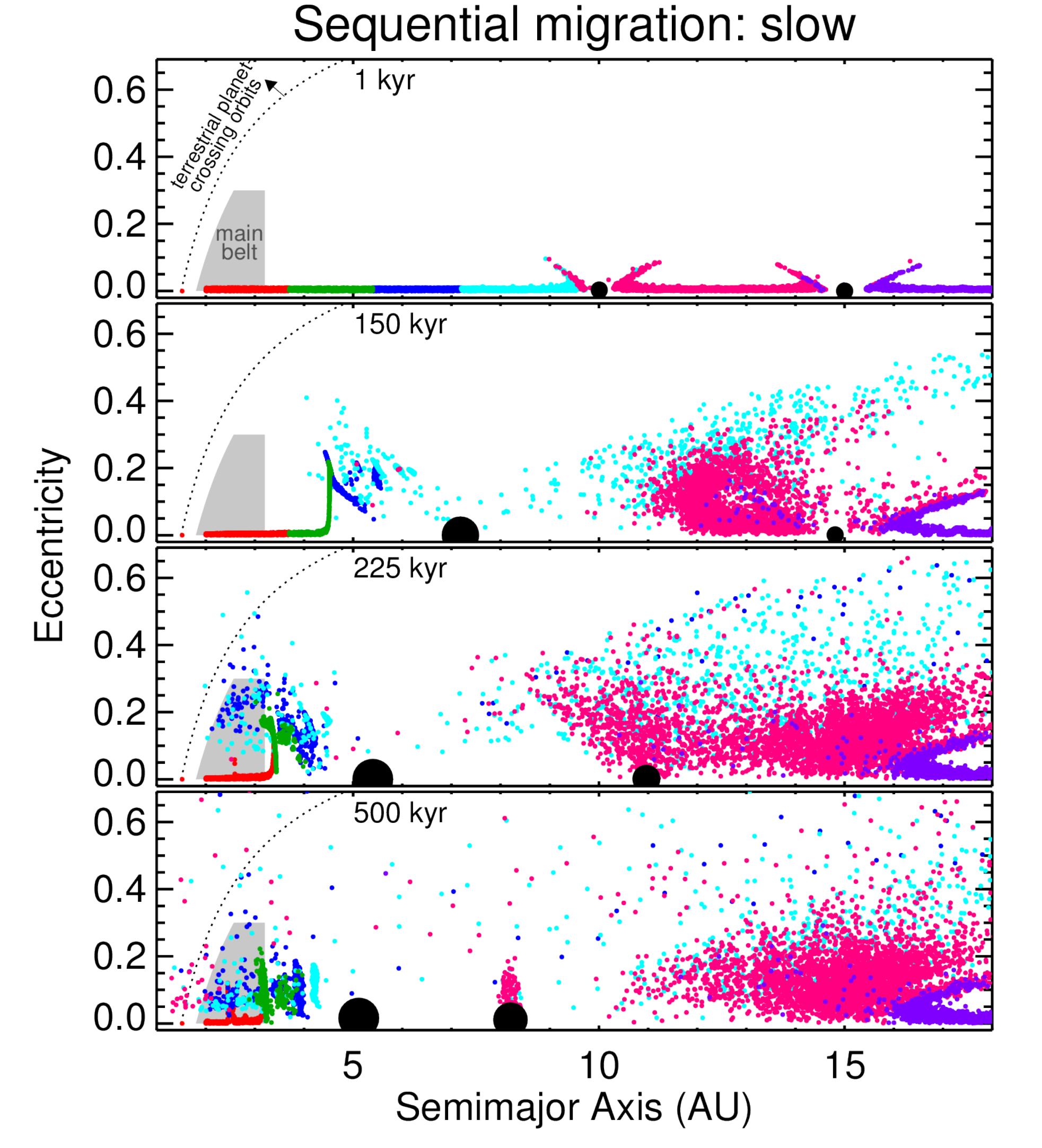}
    \caption[]{Snapshots in the evolution of four simulations with concurrent growth and orbital migration.  Here, Jupiter and Saturn's cores started at 10 and 15 AU, respectively.  In the top two panels, Jupiter and Saturn grew and migrated inward in concert, with migration and growth timescales of 100 kyr  (top left panel) and 200 kyr (top right panel).  In the bottom panels Jupiter grew first and migrated inward, followed by Saturn.  In the bottom left panel, Saturn's migration was fast enough to be captured in 2:3 mean motion resonance whereas in the bottom right panel Saturn's migration was slower and it was caught in 1:2 resonance.  In all cases the disk dissipated exponentially on a 200 kyr timescale.  Planetesimal colors track their starting position -- the four inner colors match those from Fig.~\ref{fig:jupevol}.  The vertical lines of planetesimals interior to Jupiter's orbit are strong mean motion resonances that shepherd planetesimals inward~\citep{tanaka99,zhou05,fogg05,raymond06c,mandell07}. } 
     \label{fig:migallall}
\end{center}
\end{figure}

We performed simulations with different assumptions regarding Jupiter and Saturn's migration (Figure~9). The simulations started with Jupiter's core at 10 AU and Saturn's core at 15 AU. The scenarios we tested can be divided into two categories: either Jupiter and Saturn grew and migrated together or separately.  For simultaneous migration of the two planets (top 2 panels in Fig.~9) we also tested different growth and migration rates.  For sequential migration, Jupiter first grew and migrated, followed by Saturn.  We tested different migration rates for Saturn.  As expected, fast migration led to capture into the 2:3 mean motion resonance with Jupiter, and slower migration led to capture into the 1:2 resonance~\citep{pierens08,pierens14}.  We performed simulations for planetesimal sizes of 1, 10, 100 and 1000~km, although the sims from Fig.~9 all have 100~km planetesimals. During the planets' migration, the gas disk profile was scaled and translated to keep the gaps centered on the planets' orbits and the disk's surface density continuous.

\begin{figure}
\begin{center}
\includegraphics[width=0.5\textwidth]{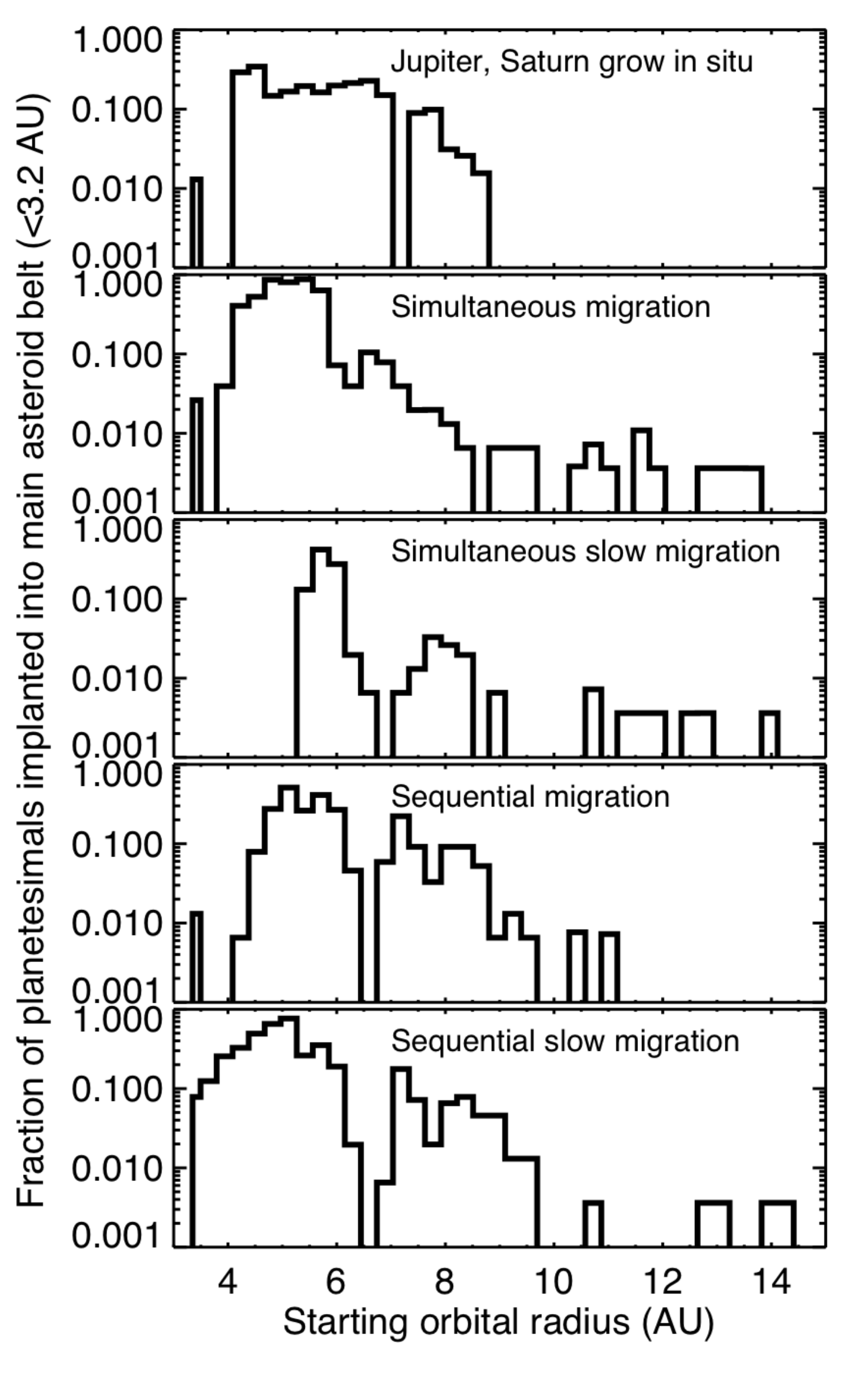}
    \caption[]{Source regions for planetesimals implanted as C-type asteroids into the main belt ($<$3.2 AU) from five different simulations.  The top panel represents the simulation from Fig.~\ref{fig:jupevol}. in which Jupiter and Saturn grew in-situ.  The other four simulations are those from Fig.~9, in which Jupiter and Saturn's cores started at 10 and 15 AU, respectively, and migrated inward.  }
         \label{fig:ctypes}
         \end{center}
\end{figure}

Figure~\ref{fig:ctypes} shows the source region of planetesimals implanted into the asteroid belt for the four simulations from Fig.~9 as well as the in-situ accretion simulation from Fig.~\ref{fig:jupevol}. The feeding zones of all five simulations have a similar shape, with the bulk of implanted C-type asteroids originating between 4-5 and 9-10 AU.  Each of the simulations with migration has a tail of implanted bodies that extends out to Saturn's starting orbital radius of 15~AU.  This source region should indeed extend out to at least the starting orbital radius of the giants' cores, which could plausibly be as distant as 25-30 AU~\citep{bitsch15}.  The simulation with slow simultaneous migration did not implant asteroids from interior to 5 AU because planetesimals were preferentially shepherded by the 2:1 mean motion resonance, which deposited a large amount of bodies just exterior to the main belt.  

The source region for implanted asteroids (Fig.~\ref{fig:ctypes}) is roughly as the one inferred from the Grand Tack model~\citep{walsh12}. We expect this to be even wider when migration of the ice giants~\citep{izidoro15b} -- which provide an additional source of planetesimals to be scattered by Jupiter -- are accounted for. The initial compositional diversity within this region may be the origin of the variety of types of asteroids in the outer belt~\citep{demeo15} as well as the variety of different volatile-rich meteorites. 

\section{Simulations including the ice giants}

We ran two additional simulations including Uranus and Neptune. In the simulations both ice giants were assumed to be fully-grown at the start~\citep[see][for a more realistic growth scenario]{izidoro15c}.  We assumed that all four giant planets migrated inward and tested both a sequential and simultaneous case.  The planets' initial orbital radii were 10, 15, 20, and 25 AU. As in the simulations from Fig.~\ref{fig:water}, the disk dispersed with $\tau_{gas} = 250$~kyr and the simulations were integrated for 2.5 Myr.  Planetesimals were 100~km in diameter.

In the sequential case, Jupiter's migration and gas accretion both took place from 100 to 400 kyr.  Saturn grew from 300 to 400 kyr, then migrated inward rapidly in just 25 kyr. Uranus migrated inward from 500 to 700 kyr and Neptune from 600 to 800 kyr. In the simultaneous migration simulation Jupiter's growth and migration again took place from 100 to 400 kyr.  Saturn grew and migrated on the same timescale.  The ice giants again migrated from 500 to 700 kyr and from 600 to 800 kyr.  The final giant planet configuration is a resonant chain with Jupiter and Saturn in 3:2 resonance, Saturn and Uranus in 2:1 resonance, and Uranus and Neptune in 4:3 resonance.

\begin{figure}
\begin{center}
\includegraphics[width=0.49\textwidth]{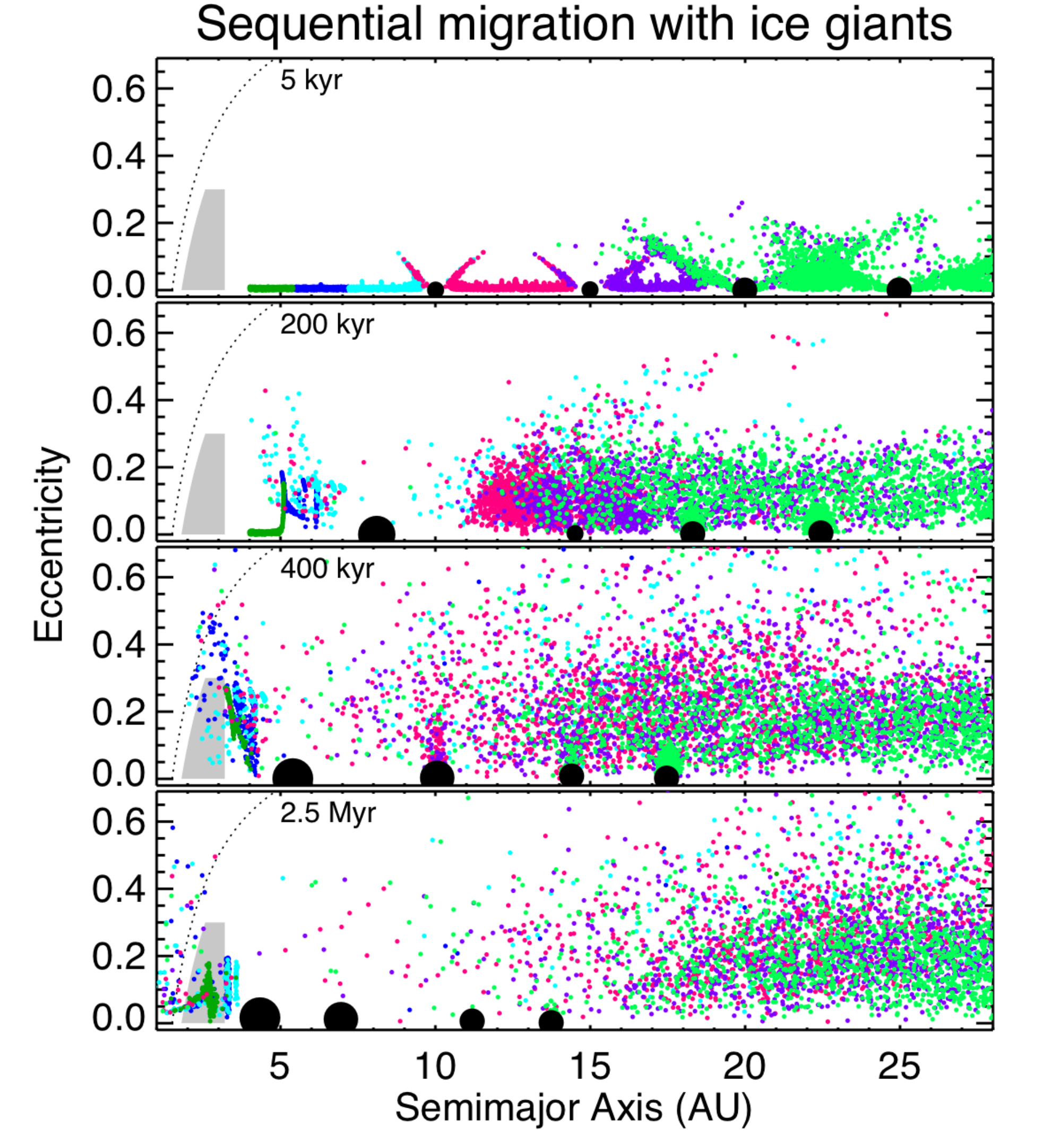}
\includegraphics[width=0.49\textwidth]{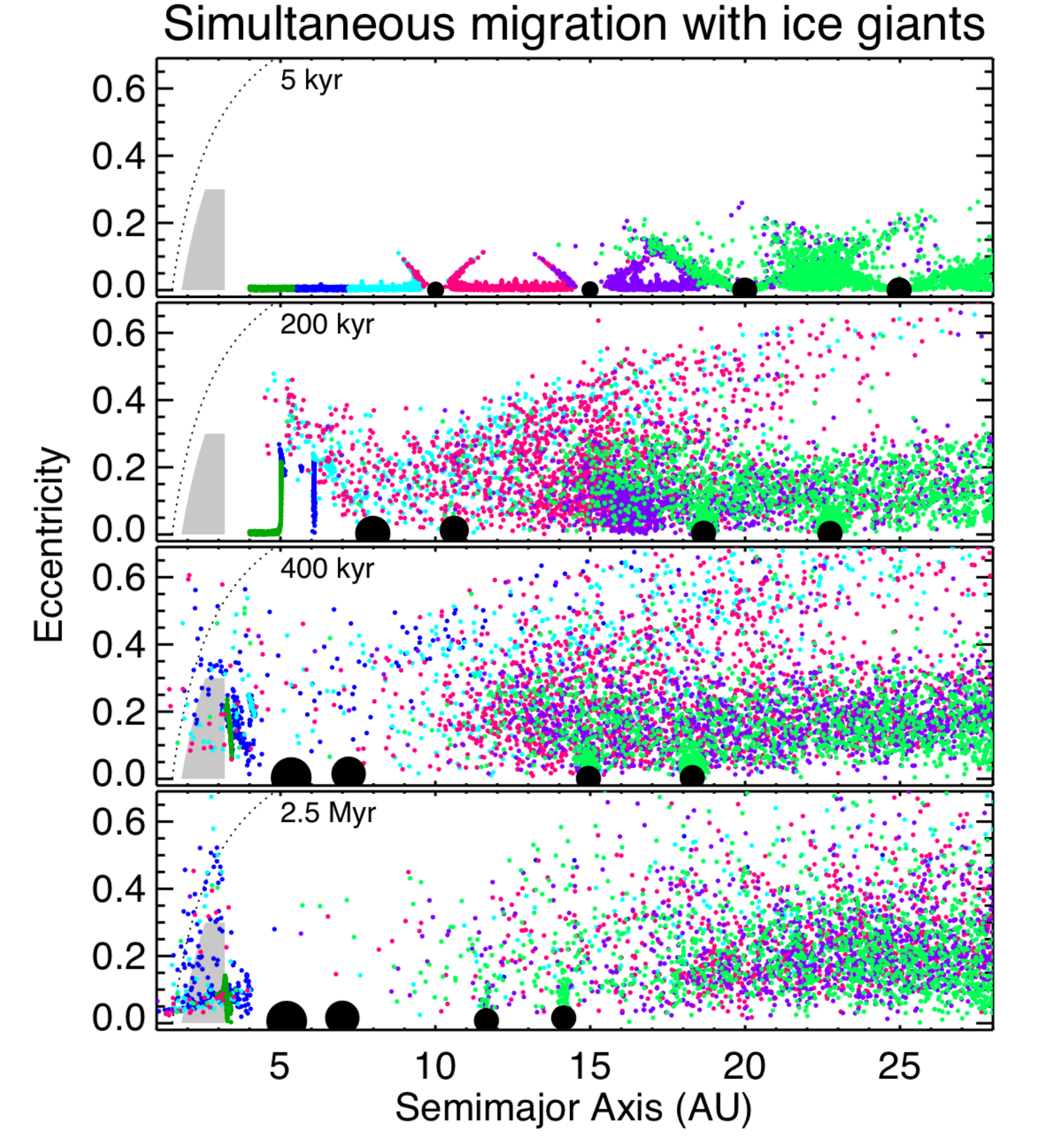}
    \caption[]{Snapshots in the evolution of two simulations with concurrent growth and orbital migration that also include the ice giants. Planetesimal colors are compatible with previous figures, with the addition of light green particles that originated past 20 AU.} 
     \label{fig:jsun}
\end{center}
\end{figure}

Figure~\ref{fig:jsun} shows the evolution of the two simulations, which included $10^4$ planetesimals initially distributed from 4 to 30 AU.  In the simulation with sequential migration, Saturn's migration gave Jupiter a substantial kick that pushed it interior to its destined final orbital radius, finishing at 4.33 AU, while Saturn stopped at 7 AU, just exterior to the 2:1 resonance.  The simulation with simultaneous migration was better behaved, with the four giant planets finishing in the expected resonant chain.  

As in Fig.~\ref{fig:migallall}, Jupiter shepherded planetesimals interior to strong mean motion resonances, and there was a prolonged phase of chaotic planetesimal scattering that both deposited planetesimals into the main belt and scattered some toward the growing terrestrial planets

\begin{figure*}
\begin{center}
\includegraphics[width=0.75\textwidth]{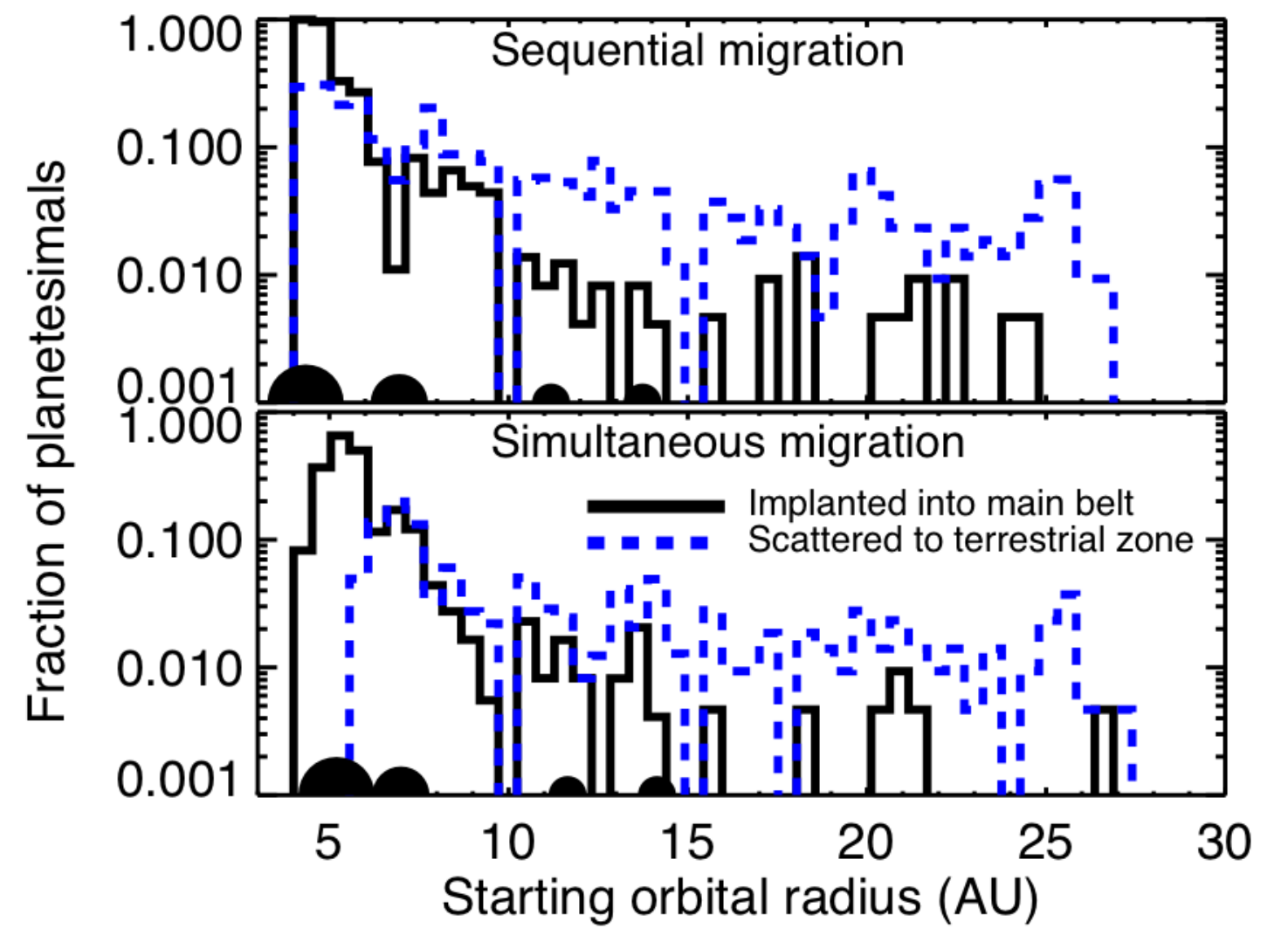}
    \caption[]{Source regions for planetesimals implanted as C-types in the main belt (black curves) and scattered to the terrestrial zone (blue dashed curves) from two simulations that include the ice giants (see Fig.~\ref{fig:jsun}).}
         \label{fig:ctypes_jsun}
         \end{center}
\end{figure*}

Figure~\ref{fig:ctypes_jsun} shows the source region for implanted and potentially water-delivering planetesimals. In simulations with ice giants particles were implanted into the main belt from a region more than 20 AU in width.  As in Fig.~\ref{fig:ctypes}, the particles with the highest efficiency of implantation were those initially located in the (present-day) Jupiter-Saturn region between roughly 4 and 10 AU.  However, the tail of the distribution of implanted planetesimals extended out to Neptune's original orbit of 25 AU.  

Planetesimals scattered toward the terrestrial planets span roughly the same radial range as the distribution of planetesimals implanted in the main belt, from 4 AU to past 25 AU (Fig.~\ref{fig:ctypes_jsun}).  However, planetesimals scattered toward the terrestrial planets have a somewhat flatter distribution and more often originate past 10 AU.  This is because planetesimals from past 10 AU only reach the inner Solar System after being scattered by multiple planets, and this tends to happen at late times, after the disk's density has already dropped substantially. Given the importance of the disk density in balancing scattering vs. implantation (Fig.~\ref{fig:water}), these late scattered planetesimals are more likely to reach the terrestrial planets than to be deposited on stable orbits within the main belt.

We have only illustrated two of a continuum of possible growth and migration pathways for the giant planets.  The mechanism of planetesimal scattering and implantation is remarkably robust.  In all cases, planetesimals from the Jupiter-Saturn region are implanted into the main belt with a high efficiency, with a tail extending out to the outermost planets. Planetesimals from across the Solar System are scattered by the growing/migrating giant planets onto high eccentricity orbits -- preferentially late in the disk lifetime -- that cross the terrestrial zone and have the potential to deliver water to the growing terrestrial planets.

\section{The high efficiency of implantation}

Implantation of scattered planetesimals into the main belt is surprisingly efficient. As shown in Fig~\ref{fig:ctypes}, a large fraction of planetesimals initially located between 4 and 10 AU are implanted into the asteroid belt during the giant planets' growth. In all cases this fraction is higher than 10\%. However, the current asteroid belt is very low in mass. There are a number of possible explanations:
\begin{itemize}
\item {\bf A large number of planetesimals were indeed implanted but the asteroid belt was subsequently depleted.}  Some disk models predict that the primordial belt contained $1-2 \mearth$ in solids~\citep{hayashi81}. An Earth-mass of C-types would indeed have been implanted into the main belt if a) there was $5-10 \mearth$ in planetesimals in the giant planet region when Jupiter underwent rapid gas accretion (Fig.~\ref{fig:imp_scat}), and b) Jupiter's gas accretion was early enough to favor asteroidal implantation (Fig.~\ref{fig:water}). This requires a mechanism to subsequently strongly deplete the belt, and there are several candidate mechanisms. Resident planetary embryos in the belt can efficiently excite and deplete the belt~\citep{wetherill92,chambers01b,petit01,obrien07}, although in many simulations embryos survive after the terrestrial planets have formed, in conflict with the observed Solar System, as the imprint of embryos on surviving planetesimals would likely survive a Nice model instability~\citep{raymond09c}. Sweeping secular resonances during the disk's dissipation are another possibility~\citep{lecar97,nagasawa05}, although it is unclear whether sweeping resonances can provide strong enough depletion~\citep{obrien07}, in particular when accounting for self-consistent (generally low-eccentricity) orbits for the giant planets. In the Grand Tack model, Jupiter's inward-then-outward migration depletes the belt by roughly two orders of magnitude~\citep{walsh11,walsh12}. Our implantation mechanism is thus consistent with (but does not require) the Grand Tack. We discuss how implantation affects the Grand Tack model in Section 8.4.

\item {\bf There were very few planetesimals nearby when the giant planets formed.}  If the giant planets formed from a swarm of planetesimals~\citep{thommes03,levison10}, it stands to reason that there would be an abundant population of planetesimals in the cores' vicinity when they undergo runaway gas accretion (albeit with orbits sculpted by the growing cores). However, this model has significant problems, in particular with regards to the inefficiency and long duration of core growth~\citep[e.g.][]{levison01}.  In contrast, the pebble accretion model proposes that giant planet cores grow preferentially by the accretion of small pebbles rather than planetesimals~\citep{lambrechts12,lambrechts14,levison15}. While some planetesimals are needed to act as `seeds' for growth by pebble accretion, it is unclear how efficiently planetesimals themselves form~\citep{johansen14}. One can envision a scenario in which a small ring of planetesimals formed, with the giant planet cores representing the fastest-growing planetesimals. Only a small population of slower-growing planetesimals was close enough to be scattered and implanted. This is plausible based on current understanding but depends on the timing and spatial extent of planetesimal formation, which is currently poorly-constrained~\citep{johansen14}. In addition, planetesimals may have been removed by collisional grinding.

\item{\bf Jupiter and Saturn formed late.}  Given that  implantation depends strongly on the gas surface density (Fig.~\ref{fig:water}), late growth of the giant planets would imply a low efficiency of implantation into the belt.  Some models suggest that Jupiter and Saturn did form late in the disk's lifetime~\citep{thommes08}. Of course, the planets cannot have formed arbitrarily late because the disk must have enough mass to supply.  The default gas disk in our simulations follows an $r^{-1}$ surface density profile normalized to $\Sigma_0 = 4000 \, g \, cm^{-2}$ at 1 AU, or roughly $0.5 \, M_J \, AU^{-2}$.  With an $r^{-1}$ surface density profile the mass in an annulus with a given width $dr$ is constant, simply $2 \pi \Sigma_0 dr$ or roughly $6 M_J$ in any given annulus $dr =$~1 AU in width.  It is therefore hard to imagine Jupiter having undergone rapid gas accretion after the disk had dissipated by more than roughly a factor of 6 (about 2 e-folding timescales), assuming that Jupiter efficiently accreted gas within an annulus of 1 AU in width~\citep[although we note that Jupiter could continue accreting gas from tidal streams for longer timescales; e.g.,][]{lissauer09}. In Fig~\ref{fig:water}, Jupiter should have formed before roughly $5 \times 10^5$ years, but it is important to realize that time zero in Fig~\ref{fig:water} is when the disk starts to disperse, likely several millions of years after CAIs~\citep{haisch01,mamajek09}. By the same arguments, Saturn should have formed before $3 \times 10^5$ years.  Pushing the formation times of each planet to late times, the inner disk would be depleted to a corresponding degree, dropping the efficiency of implantation (Fig.~\ref{fig:water}).  

\cite{kruijer17} showed that carbonaceous and non-carbonaceous meteorites originate from two co-eval but spatially separate reservoirs.  They used the meteorites' age distribution to argue that Jupiter must have been present and at least $\sim 20 \mearth$ by 1 Myr after CAIs. Jupiter's core would provide a dynamical barrier between the carbonaceous and non-carbonaceous reservoirs~\citep[see][]{lambrechts14b} and maintain their observed chemical and isotopic differences.  \cite{kruijer17} argue that Jupiter only grew to its current size 3-4 Myr after CAIs because  carbonaceous chondrites continued to form until that time~\citep{kita12,connelly12}.  Given this timeline, Jupiter's growth must certainly have triggered planetesimal scattering leading to asteroidal implantation and perhaps water delivery to the growing Earth.  However, the crucial parameter -- the surface density of the disk, in particular interior to Jupiter (Fig.~\ref{fig:water}) -- remains unconstrained.  

\item {\bf Our gas disk profile overestimates the density in the inner disk.}  The most important factor in determining the implantation efficiency is the level of orbital energy loss via gas drag in the asteroid region.  Motivated by observations~\citep{andrews09,andrews10} and viscous disk theory~\citep{lyndenbell74,chambers06,bitsch15} we chose a simple power-law profile for the disk. New models that account for non-linear hydrodynamical effects find a variety of disk profiles~\citep{bai16,suzuki16,morbyraymond16}.  One model~\citep{suzuki16} that invokes disk winds as a mechanism for angular momentum transport finds a surface density profile that peaks in the Jupiter Saturn region and drops significantly closer-in.  Such a disk would simultaneously contain enough mass in gas to grow Jupiter at 5 AU and have a low-density inner region with weak gas drag and inefficient implantation of planetesimals into the asteroid belt (and correspondingly more efficient delivery of water-bearing planetesimals to the terrestrial planet region).   
\end{itemize}

At present it is unclear which of these solutions is the most reasonable.  The field of planet formation is presently making significant strides in learning about pebble accretion, planetesimal formation, giant planet formation, and disk models so we expect this to become better understood in the years to come.

\section{Matching the inner Solar System}

We now discuss how our results can match constraints and fit within the broader context of Solar System formation. We discuss constraints from the relative abundance of water in C-types vs. the Earth (Section 8.1), isotopic constraints from Solar System bodies (Section 8.2), and the structure of the asteroid belt (Section 8.3).  We then show how our mechanism fits within different models of terrestrial planet formation (Section 8.4).  

\subsection{Relative abundance of C-types and Earth's water}

Earth's bulk water content is debated, as the amount of water trapped in the mantle is poorly-constrained.  Estimates range from less than one ocean~\citep{panero16} to a few~\citep{halliday13} to more than ten oceans~\citep{marty12}, where an ``ocean'' of water is defined as the amount of water on Earth's surface ($\approx 1.5 \times 10^{24} \, g$). The core's water (or hydrogen) content is likewise debated~\citep{badro14,nomura14}.

Planetesimals scattered onto inner Solar System-crossing orbits represent a source of water for the terrestrial planets. These objects originated beyond 4 AU and their source region overlaps with that of planetesimals implanted as C-type asteroids (see Figs.~\ref{fig:imp_scat} and ~\ref{fig:ctypes}). Given the chemical match between as Earth's water and Nitrogen have the same isotopic signature as carbonaceous chondrite meteorites~\citep{marty06}, linked with C-types (see Section 8.2 below). 

Earth's water content is consistent with the accretion of $2.5-25 \times 10^{-3} \mearth$ of C-type material~\citep{morby00,marty12}. The total present-day mass in C-types is $\sim5 \times 10^{-4} \mearth$. Accounting for a factor of $\sim$5 in dynamical depletion over the belt's history~\citep{minton10,morby10}, roughly 10-20 times more C-type material must have been scattered onto terrestrial planet-crossing orbits than was implanted into the asteroid belt. If a large amount of collision depletion took place after implantation~\citep[e.g.][]{bottke05b}, it would reduce this ratio.  For example, if the belt was collisionally depleted by a factor of 20, then a roughly equal amount of planetesimals would need to have been scattered to the terrestrial zone and implanted in the main belt.

A single relatively late burst of giant planet formation could in principle simultaneously implant the C-type asteroids and deliver water in the correct ratio (see Fig.~\ref{fig:water}).  However, we find it more consistent to envision multiple episodes of growth and planetesimal scattering. Jupiter likely formed while the disk was still massive~\citep{lissauer09}, implanting the bulk of the C-types. Saturn formed later and its scattered planetesimals were divided between implantation and water delivery. There were certainly later bursts of planetesimal scattering associated with the migration and growth of the ice giants~\citep{izidoro15b}, contributing mainly to terrestrial water delivery with less asteroidal implantation (see Section 6). This multiple-scattering scenario naturally predicts the observed diversity of asteroid types within the outer belt~\citep{gradie82,bus02,demeo13,demeo14}.  Of course, Saturn need not have formed after Jupiter to match the inner Solar System.  Many variants on the multiple-scattering scenario could provide a match, and these can be explored using the timing constraints illustrated in Fig.~\ref{fig:water}.

It is also worth considering the chemical constraints on Earth's accretion. \cite{rubie15} showed that matching Earth's core mass and chemistry requires that its early accretion was dominated by reduced material and that any oxidized material -- including water -- was accreted late. In this paper we do not simulate terrestrial planet formation or the actual delivery of water-rich material to the growing Earth. However, we can use previous work to make an educated guess about how our results fit in.  By the time of Jupiter's gas accretion, the terrestrial region is thought to have been made up of roughly Mars-sized planetary embryos~\citep{kokubo00,morby15a} that presumably accreted predominantly from local, reduced material. In our simulations, planetesimals scattered by the growing giant planets toward the terrestrial planet region are stranded on high-eccentricity, inclined orbits when the disk dissipates.  This is similar to the tail of water-delivering material produced by Jupiter and Saturn's late outward migration in the Grand Tack model~\citep{walsh11}, and we expect its accretion to proceed in a similar fashion.  \cite{obrien14} showed that water delivery to Earth from a distribution of scattered C-type bodies happens preferentially late in Earth's growth, but most water is accreted before the last giant impact on Earth.  Assuming a similar outcome for our simulations, the mechanism presented here appears consistent with the geochemical constraints from \cite{rubie15}.

\subsection{Isotopic constraints on source regions of water}

The strongest discriminant to date between sources of Earth's water is their isotopic ratios. Earth's hydrogen and nitrogen isotopic signatures (the D/H and $^{15}N/^{14}N$ ratios) match carbonaceous chondrite meteorites~\citep{marty06}, spectroscopically linked with C-type asteroids~\citep{bus02}. Water directly accreted from nebular gas would have had a much lower D/H ratio than the Earth's present-day value, and some low D/H water has indeed been found from deep mantle sources~\citep{hallis15}. Comets appear to have a broad range of D/H values, from Earth-like~\citep{hartogh11,lis13} to several times higher~\citep{altwegg15}.  However, their $^{15}N/^{14}N$ ratios are systematically higher than Earth's~\citep{marty06}. Assuming that Comet 67P/Churyumov-Gerasimenko is representative of the bulk cometary reservoir, \cite{marty16} used ROSINA measurements from the ROSETTA mission to show that comets represent a contribution of less than 1\% for Earth's water (although they may have provided the bulk of certain atmospheric noble gases).  Carbonaceous chondrites -- particular CI chondrites -- appear to represent the best match to Earth's water and nitrogen~\citep{alexander12}. 

Can an object's D/H ratio be used to determine where it condensed?  It is generally expected that the D/H increased with orbital radius within the Sun's planet-forming disk~\citep{drouart99,jacquet13}.  However, models disagree on the radial behavior of the D/H ratio in the disk. \cite{yang13} found a high D/H in the Jupiter-Saturn region but a lower D/H at larger distances. In contrast, \cite{albertsson14} found a D/H ratio that maintained Earth-like values out to $\sim$2.5 AU for a laminar disk but out to $\sim 10$~AU for a turbulent disk.

\cite{alexander12} pointed out that Enceladus' measured D/H ratio is twice as high as Earth's, and suggested that Earth's water must therefore have its origins interior to Saturn.  The question is whether this data point indicates that all water from beyond Saturn has a high D/H and is thus incompatible with being the primary source of Earth's water. There are many questions to be asked. Is Enceladus' D/H a constraint for the global planet-forming disk's D/H, or for the local conditions in the much higher-pressure moon-forming disk around the young Saturn?  Where did Saturn's core originate in relation to the condensation location of different classes of comets?  And how does Titan's Earth-like D/H ratio \cite[measured in CH$_4$;][]{abbas10} fit within this picture?

We have shown that it is inevitable that planetesimals from near Saturn were implanted into the asteroid belt, likely as C-types. This implanted population had the same source region as Earth's water (Fig.~\ref{fig:imp_scat}). Although there may be some differences in the timing of scattering vs. implantation, we naively expect their isotopic ratios to be similar. It remains to be understood whether this isotopic variety can be mapped back to the original condensation region or time~\citep[see][]{krot15}.

We conclude that our implantation mechanism appears to be consistent with current constraints.

\subsection{The asteroid belt's orbital structure and size distribution}

Other constraints on the early evolution of the asteroid belt come form the belt's orbital structure and size distribution.  We briefly discuss these constraints, arguing that our implantation mechanism is consistent with each.

While low in total mass the asteroids' orbits are excited in eccentricity and inclination. Several potential sources of excitation have been proposed, including excitation by local planetary embryos~\citep{wetherill92,chambers01b,petit01,obrien07}, sweeping secular resonances~\citep{lecar97,zheng17}, and chaotic excitation by Jupiter and Saturn~\citep{izidoro16}.  In our simulations the implanted C-types have size-dependent eccentricities and inclinations (Fig.~\ref{fig:long_2}) because of the size-dependent strength of aerodynamic gas drag~\citep{adachi76}. In a high-density or long-lived disk, all planetesimals' eccentricities and inclinations are damped to near zero. In contrast, in a lower-density disk, only the smallest planetesimals have their eccentricities damped to low values and there is a natural spread. If Jupiter and Saturn formed relatively late -- or if the inner disk's surface density were lower than in our simulations -- then we can imagine a much broader distribution of eccentricity and inclination among implanted asteroids. Of course, later phases of dynamical perturbation from secular resonance sweeping~\citep{lecar97,nagasawa00,zheng17}, chaotic excitation~\citep{izidoro16}, an instability in the giant planets' orbits~\citep{morby10}, and long-term dynamical effects~\citep{minton10} may have contributed to shaping the present-day distribution. We note, however, that \cite{obrien07} found that secular resonance sweeping in a realistic context (for reasonable disk dissipation timescales) cannot match the belt's depletion or orbital structure.

The asteroid belt's size distribution constrains the early collisional evolution of the belt~\citep{bottke05}.  Several different initial asteroid size distributions have been proposed~\citep{morby09,weidenschilling11,johansen15}. Our mechanism of planetesimal implantation is modestly size-dependent~\citep[as is the secular resonance sweeping model of][]{zheng17}.  Mid-sized asteroids (10-100~km) are the most efficiently implanted across the belt (Figs.~\ref{fig:jsdistr} and~\ref{fig:imp_scat}). While the observed belt does show a deficit of very large and very small asteroids~\citep{bottke05}, we cannot claim to match the present-day size distribution given the unconstrained size distribution of the implanting population. Yet the fact that our mechanism implants a broad range of sizes suggests that it is not at odds with the observed size distribution.


\subsection {Our results in the context of terrestrial planet formation models}

This paper introduces a new physical mechanism: scattering of planetesimals by growing gas giant planets leading to implantation in the asteroid belt or water delivery to growing terrestrial planets.  This is a {\em mechanism}, in that this naturally occurs anytime a giant planet forms, rather than a {\em model} designed to reproduce the Solar System. In this section we discuss how the mechanism fits within four models of planet formation in the inner Solar System: the classical model, the Grand Tack model, the low-mass asteroid belt model, and the pebble accretion model. 

{\bf The classical model}~\citep{wetherill96,chambers98,chambers01,levison03,obrien06,raymond06b,raymond09c,raymond14,morishima08,morishima10,fischer14} proposes that the giant planets formed near their current locations and influenced the late stages of terrestrial planet accretion. In the classical model, the terrestrial planets' feeding zones extended into the asteroid region, and material originating past 2.5-3 AU represents the source of the terrestrial planets' water~\citep{morby00,raymond04,raymond07a,obrien06}.  The initial distribution of water-bearing in the classical model material reflects the present-day distribution of primitive classes of asteroids, with water-rich material existing only beyond 2.5-2.7 AU with C-type compositions~\citep{abe00}.  Given the reliance on the distribution of asteroid classes, we can simply interpret this distribution of water-rich material as having been shaped by planetesimal scattering and implantation at an earlier epoch when the giant planets formed.  Planetesimal scattering would also naturally have delivered water-rich material inward to the terrestrial planets, making that a weaker constraint on formation models.  Of course, the classical model is currently disfavored.  By assuming ordered accretion, the classical model tends to produce planetary systems in which neighboring planets have similar masses~\citep{lissauer87,kokubo00}. The model fails to match the large Earth/Mars and Venus/Mercury mass ratios~\citep{wetherill91,raymond09c} in all but a few percent of simulations~\citep{fischer14}. 

{\bf The Grand Tack model}~\citep{walsh11,raymond14c,obrien14,jacobson14,brasser16,brasser17} invokes the orbital migration of Jupiter to sculpt the inner Solar System. In the Grand Tack model, Jupiter grew large, carved a gap in the disk and migrated inward. Saturn formed later, migrated inward and caught up with Jupiter when Jupiter was at ~1.5-2 AU.  At this point the direction of migration was reversed~\citep{masset01,morby07a,pierens08} and the two planets migrated back outward until the disk dispersed~\citep{pierens11}. JupiterÕs inward migration depleted the asteroid belt and the Mars region but not the Earth-Venus zone.  In the Grand Tack model, S-type asteroids were scattered outward during Jupiter's inward migration, then back inward during its outward migration and the C-types were implanted from exterior orbits during outward migration. As the Grand Tack model starts with a fully-formed Jupiter, it neglects an earlier phase of planetesimal scattering.  Given that the disk was relatively dense at this time and that in the model Jupiter was closer-in (at $\sim 3.5$~AU), a large fraction of nearby planetesimals would have been deposited into the terrestrial planet region during this scattering. The division between S-type objects interior to Jupiter's initial orbit and C-type objects exterior to Jupiter is overly simplistic.  Some water would have delivered to the terrestrial planets at an earlier time than currently calculated in the model~\citep{obrien14}, in the form of planetesimals scattered inward during Jupiter's rapid gas accretion and shepherded inward during Jupiter's migration. Of course, the extent of early pollution depends on the initial distribution of planetesimals, which is unconstrained. 

{\bf The low-mass asteroid belt model}~\citep{drazkowska16,izidoro15c,izidoro16,morbyraymond16} proposes that the asteroid belt never contained a large mass in planetesimals. One model of dust drifting in evolving planet-forming disks found a pileup capable of producing planetesimals in a narrow region near 1 AU~\citep{drazkowska16}. This confined annulus of material can match the large Earth/Mars mass ratio~\citep{hansen09,walsh16}. In the low-mass asteroid belt model, the asteroid belt was populated by a modest population of native asteroids as well as by bodies scattered outward from the terrestrial planet region~\citep{bottke06}. In a companion paper we show that even if the primordial asteroid belt was completely empty it would have been populated by S-types from the terrestrial planet region~\citep{raymond17b}. Our results fill an obvious hole in the model by providing a source of C-type asteroids, implanted during Jupiter growth.  Of course, since the belt is initially low in mass and no later depletion of the belt is needed, only a small mass in C-types would have been implanted.  As discussed above, there are a number of solutions related to the initial planetesimal distribution, the timing of giant planet formation and the disk's profile.

{\bf Pebble accretion.} It has been proposed that the terrestrial planets may have grown in large part by accreting pebbles drifting inward through the disk. Simulations including pebble accretion appear to match the Earth/Mars mass ratio, at least in the context of non-migrating giant planets~\citep{levison15b}. While this model warrants further study, it will certainly be affected by our planetesimal scattering mechanism.  While Jupiter's gas accretion creates a pressure bump that traps small particles and shuts off the flux of pebbles into the inner Solar System~\citep{lambrechts14b}, it also delivers a population of planetesimals that match the C-types and deliver water to the growing terrestrial planets.  

To summarize, the mechanism of planetesimal scattering can be incorporated into each of these models of terrestrial planet formation in a coherent way.

\subsection{Comparison with previous results}

We are not aware of previous work with the same approach or results as our paper.  However, three studies are worth mentioning.

First, \cite{grazier14} performed simulations of planetesimal scattering by the giant planet cores in the absence of gas drag.  They found that a large fraction of planetesimal from the giant planet region were scattered onto orbits that crossed the outer asteroid belt, and interpreted this as pollution of the outer asteroid belt.  While our results agree in principle, we find that little material is scattered to the asteroid belt by Jupiter and Saturn before they undergo rapid gas accretion because they cannot scatter planetesimals strongly enough to overcome the effects of gas drag and reach the main belt (Fig~\ref{fig:jupevol}). 

Second, \cite{turrini12} simulated the effect of Jupiter's growth on the orbital evolution of the asteroid belt~\citep[see also][]{turrini14}.  They found a strong increase in collision velocities associated with Jupiter's growth, as well as radial mixing between planetesimals originating in different zones.  They did not include gas drag in their simulations so we cannot compare capture rates, nor do we focus on collision velocities.  Yet the results seem broadly consistent with Jupiter's growth having a strong effect on the belt.  
 
We wonder whether the strong increase in collision velocity at the time of Jupiter's growth seen by \cite{turrini12} could be associated with the large impact velocity needed to explain the CB chondrites~\cite{krot05}.  CB chondrites are dated to $\sim 4.8$~Myr after CAIs~\citep{bollard15}. \cite{johnson16} associated the spike in impact velocity with the timing of Jupiter's migration (assuming a Grand Tack), but an alternate explanation is that it simply measures the timing of Jupiter's rapid gas accretion.

Third, \cite{brasser07} studied the formation of the Oort cloud from planetesimals scattered by the giant planets in the presence of the gaseous disk. They found that the smallest comets were too strongly-coupled to the gaseous disk to be scattered out to the Oort cloud and were instead trapped in exterior resonance with giant planets (assumed to be fully-formed) or in some cases drifted inward into the asteroid belt after being scattered by Jupiter.  This is similar to the evolution of the $D=1$~km planetesimals in our simulations.  \cite{brasser07} found that larger ($D \gtrsim 20$~km) comets were decoupled enough from the gas to be scattered onto high-eccentricity orbits that could reach the Oort cloud.  Again, this is similar to our results in that gas drag causing size-sorting of planetesimals.

\section{Conclusions}

We have shown that a giant planet's growth induces large-scale radial mixing of small bodies. Planetesimals destabilized by Jupiter and Saturn's growth were scattered in all directions, and a substantial fraction were trapped in the outer parts of the main asteroid belt. By identifying the precursors of C-types with objects that condensed between roughly 4 and 9 AU, this naturally explains the distribution of C-types within the main belt.  Of course, the source region of implanted asteroids was likely even wider, extending out to $\sim 15$~AU or beyond when migration is accounted for.

The same mechanism delivered water to the growing terrestrial planets. In the limit of weak gas drag -- corresponding to large planetesimals or a lower-density gas disk -- Jupiter scatters planetesimals {\it past} the asteroid belt onto orbits that cross the terrestrial planet region (Fig.~\ref{fig:water}). Planetesimals scattered at early times, such as when Jupiter is thought to have formed, are more likely to be trapped as asteroids.  Later planetesimal scattering events -- triggered by the growth and/or migration of Saturn and the ice giants -- deliver water to the terrestrial planets with less efficient implantation. This mechanism appears to be consistent with Solar System constraints (Section 7). 

On a philosophical level, our results imply that Earth's water and the C-type asteroids are a simple, unavoidable consequence of the giant planets' growth.  

To conclude, we stress that any time a giant planet forms it scatters planetesimals onto closer-in orbits. As most giant exoplanets are thought to have formed beyond the snow line~\citep{alexanderpascucci12,bitsch15b}, this implies large-scale inward transport of water-rich bodies. Giant exoplanets therefore likely to have delivered some water to the inner regions of their host systems.

\vskip .5in
{\it Acknowledgments.} 
We thank Ramon Brasser and two anonymous referees for their constructive reports. We acknowledge discussions with A. Mandell, A. Morbidelli, A. Pierens, S. Jacobson, R. Gomes, T. Guillot and M. Lambrechts. We thank the Agence Nationale pour la Recherche for support via grant ANR-13-BS05-0003-002 (grant MOJO).  A. I. thanks financial support from FAPESP (Process numbers: 16/12686-2 and 16/19556-7).  S.~N.~R. is also grateful to NASA Astrobiology InstituteÕs Virtual Planetary Laboratory Lead Team, funded through the NASA Astrobiology Institute under solicitation NNH12ZDA002C and Cooperative Agreement Number NNA13AA93A.

A historical note: S.N.R. initiated this project in 2008. It was planned as a ``prequel'' to the Grand Tack model (led by S.N.R. at the time).  A bug in the code led to the project being postponed, and its importance overshadowed by the publication of the Grand Tack model. In 2015, A.I. resolved the bug, upgraded the code as a whole, and made this paper possible.


\end{document}